\documentclass[a4paper,journal]{IEEEtran}
 \usepackage[utf8]{inputenc}
 \usepackage{amsmath,amssymb,amsfonts,bm}
 \usepackage{algorithmic}
 \usepackage{algorithm}
 \usepackage{array}
 \usepackage{textcomp}
 \usepackage{stfloats}
 \usepackage{url}
 \usepackage{verbatim}
 \usepackage{graphicx}
 \usepackage{cite}
 \usepackage[caption=false,font=footnotesize]{subfig}
 
 \usepackage{color}
 \usepackage[acronyms]{glossaries}

\usepackage{pgfplots}
\pgfplotsset{compat=newest}
\usepackage{tikz}
\usetikzlibrary{calc}
\usetikzlibrary{fit}
\usetikzlibrary{positioning}
\usetikzlibrary{arrows.meta}

\usepackage[hidelinks]{hyperref}

\newcommand{\transpose}{\text{T}}
\newcommand{\hermitian}{\text{H}}

\renewcommand{\Re}{\text{Re}}

\newacronym{elbo}{ELBO}{evidence lower bound}
\newacronym{sbl}{SBL}{sparse Bayesian learning}
\newacronym{map}{MAP}{maximum a-posteriori}
\newacronym{ml}{ML}{maximum likelihood}
\newacronym{kl}{KL}{Kullbach-Leibler}
\newacronym{pdf}{PDF}{probability density function}
\newacronym{vmd}{VMD}{variational mode decomposition}
\newacronym{lse}{LSE}{line spectral estimation}
\newacronym{snr}{SNR}{signal-to-noise ratio}
\newacronym{em}{EM}{expectation-maximization}
\newacronym{roc}{ROC}{receiver operating curve}
\newacronym{awgn}{AWGN}{additive white Gaussian noise}
\newacronym{ospa}{OSPA}{optimal subpattern assignment}

\newcommand{\ist}{\hspace*{.3mm}}
\newcommand{\rmv}{\hspace*{-.3mm}}
\newcommand{\iist}{\hspace*{1mm}}

\newcommand{\nn}{\nonumber}

\begin{document}

\title{\huge Variational Inference of Structured Line Spectra Exploiting Group-Sparsity}

\author{Jakob Möderl, Franz Pernkopf, Klaus Witrisal, Erik Leitinger \vspace*{-5mm} 
\thanks{This research was partly funded by the Austrian Research
		PromotionAgency (FFG) within the project SEAMAL Front (project
		number: 880598). Furthermore, the financial support by the Christian Doppler Research Association, the Austrian Federal Ministry for Digital and Economic Affairs and the National Foundation for Research, Technology and Development is gratefully acknowledged.
		
		The authors are with the Signal Processing and Speech Communications Laboratory at Graz University of Technology, Graz, Austria. Klaus Witrisal and Erik Leitinger are further associated with the Christian Doppler Laboratory for Location-aware Electronic Systems.
	}%
}

\maketitle

\begin{abstract}
	In this paper, we present a variational inference algorithm that decomposes a signal into multiple groups of related spectral lines. 
	The spectral lines in each group are associated with a group parameter common to all spectral lines within the group. The proposed algorithm jointly estimates the group parameters, the number of spetral lines within a group, and the number of groups exploiting a Bernoulli-Gamma-Gaussian hierarchical prior model which promotes sparse solutions. Aiming to maximize the \gls{elbo}, variational inference provides analytic approximations of the posterior \glspl{pdf} and also gives estimates of the additional model parameters such as the measurement noise variance. While the activation variables of the groups and the associated group parameters (such as fundamental frequencies and the corresponding higher order harmonics) are estimated as point estimates, the remaining parameters such as the complex amplitudes of the spectral lines and their precision parameters are estimated as approximate posterior \glspl{pdf}.
	
	We demonstrate the versatility and performance of the proposed algorithm on three different inference problems. In particular, the proposed algorithm is applied to the multi-pitch estimation problem, the radar signal-based extended object estimation problem, and \gls{vmd} using synthetic measurements and to real multi-pitch estimation problem using the Bach-10 dataset. The results show that the proposed algorithm outperforms state-of-the-art model-based and pre-trained algorithms on all three inference problems.
	
\end{abstract}
\glsresetall

\begin{IEEEkeywords}
	line spectral estimation, sparse Bayesian learning, multi-pitch estimation, extended object detection, variational mode decomposition
\end{IEEEkeywords}

\section{Introduction}
\label{sec:introduction}

The problem of \gls{lse} \cite{stoica2005:SpectralAnalysis}, i.e. estimating the frequencies and amplitudes of a superposition of complex exponential functions from noisy measurements, is ubiquitous in signal processing.
Solutions to this problem are applicable in many areas of physics and engineering, including range and direction estimation in radar and sonar, speech and music analysis, wireless channel estimation, molecular dynamics and geophysical exploration.
Furthermore, in many applications the spectral lines can be organized into groups which share underlying parameters.
One such example is pitch estimation in speech or music analysis \cite{benents2019SPM,mueller2011STSP:musicAnalysis,christensen2009:MultiPitchEst,christensen2008SP}. The signal of each speaker during voiced speech or each tone of an instrument exhibits a harmonic structure with spectral lines at integer multiples of some base frequency.
Another example in radar signal processing are extended objects, which give rise to multiple related target signals \cite{granstrom2017JAIF:EO}. Transformed into the frequency domain, this results in multiple correlated lines \cite{schubert2013TAP}.
Many other problems such as \gls{vmd} \cite{dragomiretskiy2014TSP:VMD} can be approximated by a structured line spectrum.
In these examples, the number of groups as well as which spectral line belongs to which group (i.e. the group structure) is not known a priori and has to be estimated as well, further complicating the estimation process.

\subsection{State of the Art}
Common solutions to the \gls{lse} problem assume the number of spectral lines (i.e. the model order) is known and no relation exists between the spectral lines. Such examples include subspace based methods such as MUSIC \cite{schmidt1986TAP} or ESPRIT \cite{roy1989TASSP} as well as the \gls{ml} method \cite{ziskind1988TASSP:ML,feder1988TASSP:ML}.
If the model order is not known, a criterion such as the Bayesian information criterion (BIC) or the Akaike information criterion (AIC) can be used to select a model order from a set of candidate model orders \cite{stoica2004SPM}.
However, this approach can be computationally expensive since a solution must be obtained for each considered model order before a particular solution is chosen.

Sparse signal reconstruction methods aim to reconstruct a signal based on a large dictionary matrix which is weighted with a sparse amplitude vector.
Thus, the model order is estimated as part of the process, alleviating the issue.
A prominent instance of dictionary based sparse signal reconstruction method is the least absolute shrinkage and selection operator (LASSO) \cite{tibshirani1996RSSB}, which is also called basis pursuit denoising \cite{chen2001SIAMRev}. Further methods include matching pursuit \cite{mallat1993TSP:MP}, \gls{sbl} \cite{tipping1999NeurIPS:RelevanceVector,tipping2003WAIS:FastMarginalSparseBayesian, shutin2011TSP:fastVSBL} and SPICE \cite{stoica2011TSP:SPICE}.
See \cite{wipf2004TSP,wipf2011TIP} for a detailed discussion about the similarities and differences of some of these methods.
Many of these algorithms have been extended to include a group structure, such as the group-LASSO \cite{yuan2006RSSB,kyung2010BA,raman2009ICML,xu2015BA}, blockwise sparse regression \cite{kim2006SS:BSR}, block matching pursuit \cite{eldar2010TSP:BlockSparseMP}, group-\gls{sbl} \cite{zhang2011STSP:bSBL,zhang2013TSP:BlockSparseSBL}, pattern-coupled \gls{sbl} \cite{fang2015TSP:PatternCoupledSBL} and group-SPICE \cite{kronvall2017SP:groupSparseRegression}.
A disadvantage of using a fixed dictionary matrix is the spectral leakage induced by the model mismatch, which decreases the estimation performance \cite{chi2011TSP,duarte2013ACHA}.
Thus, parametrized approaches have been developed such as the gridless-SPICE algorithm \cite{yang2015TSP:GLS} and extensions of \gls{sbl} to a continuous (i.e., infinite) dictionary matrix with super-resolution capability
\footnote{We define super-resolution as the ability of an algorithm to resolve spectral lines even if their separation in the dispersion domain is below the intrinsic resolution of the measurement equipment exploiting continuous dictionary matrices.} \cite{hansen2018TSP:SuperFastLSE, hansen2014SAM:SBL, shutin2013:VSBL}.
A further development of \gls{sbl}-based super-resolution methods specific to \gls{lse} is the VALSE algorithm \cite{badiu2017TSP:VSBL}, which estimates posterior distributions of the frequencies instead of point estimates.
Note, that all sparse signal reconstruction methods with complex amplitudes can be reframed as a grouping approach, where each group consists of the real and imaginary part of each weight \cite{pedersen2015SP}.

Methods to solve the \gls{lse} problem using a grouped approach can be found for the application of multi-pitch estimation.
A few examples include a harmonic extension for the capon beamformer and the MUSIC principle, as well as an \gls{em}-based estimator, see \cite{christensen2009:MultiPitchEst} for a collection of these methods.
A more recent approach is based on block sparsity given a grid of fundamental frequencies \cite{adalbjoernsson2015SP}.
However, since this approach is based on a fixed frequency grid, it suffers the same drawbacks as other sparse signal reconstruction methods with fixed dictionary matrices.
To alleviate this issue, \cite{swaerd2018TASLP} proposes a block-sparse method for harmonic \gls{lse} based on a grouped continuous (infinite) dictionary matrix.
Finally, \cite{vincent2008NC:BayesianHarmonic} uses a Bayesian hierarchical model and proposes an adaptive factorization of the posterior.
Contrary to this work, \cite{vincent2008NC:BayesianHarmonic} is not explicitly based on sparsity.

\subsection{Contribution}

In this paper, we propose a variational inference algorithm that promotes group-sparsity by exploiting a hierarchical Bernoulli-Gamma-Gaussian model for structured line spectra.
The proposed algorithm decomposes the signal into several groups of related spectral lines which share a common group parameter. Each common group parameter is expressed by a continuous (infinite) dictionary and each spectral line within each group is related to this common group parameter by a discrete (finite) dictionary. An example for such a structured line spectrum can be a mixture of harmonic signals, where each common group parameter represents the fundamental frequency and the lines within the group form a harmonic series of spectral lines at multiples of the fundamental frequency. The contributions of this work are as follows.
\begin{itemize}
	\item We apply a layered hierarchical Bernoulli-Gamma-Gaussian model, combining the Bernoulli-Gaussian model of \cite{badiu2017TSP:VSBL} with the Gamma-Gaussian model as it is usually used in \gls{sbl} \cite{tipping2003WAIS:FastMarginalSparseBayesian} to obtain a solution which is sparse on two levels: the number of groups and the number of spectral lines within each group. The number of groups as well as the size of each group are estimated jointly with the continuous and discrete dictionary parameters.
	\item We present a formulation that allows to consider different structural relations between the spectral lines in the model. Thus, the model can be applied to a variety of inference problems.
	\item We derive the relation between the threshold governing the sparsity of groups and the threshold governing the sparsity of spectral lines within a group. This simplifies the process of tuning these thresholds to the application at hand.
	\item We demonstrate performance advantages on three different inference problems---multi-pitch estimation, detecting and estimating extended objects using radar signals and \gls{vmd}---using simulated data.
	\item We investigate the performance of the proposed algorithm on real multi-pitch data by applying it to the publicly available Bach-10 dataset.
\end{itemize}

\section{Signal Model and Bayesian Formulation}
\label{sec:signal-model}
\subsection{Signal Model}

We consider an $N$-length signal vector $\bm{x} =[x(-\frac{N}{2}T_\text{s})$  $x((-\frac{N}{2}+1)T_{\text{s}})\, \cdots\, x((\frac{N}{2}-1)T_{\text{s}})]^\transpose \in \mathbb{C}^{N}$, which contains the values of some continuous function $x(t)$ sampled at instances $\bm{t} = [-\frac{N}{2}T_{\text{s}}\iist\ist (-\frac{N}{2}+1)T_{\text{s}}\,\cdots\,(\frac{N}{2}-1)T_{\text{s}}]^\transpose$ with regular sampling interval $T_{\text{s}}$.
We assume that $\bm{x}$ is a linear combination of spectral lines in noise, and the spectral lines can be structured into $K$ groups as
\footnote{As an illustrative example for a structured line spectrum consider a note with pitch $f_0$ played on an instrument. The line spectrum of the audio signal produced by the instrument is a harmonic series with spectral lines at multiples of $f_0$, e.g. at $\{f_0,\, 2 f_0$,\, $3 f_0$,\, $4 f_0,\, 5 f_0\}$. We can model such a line spectrum using \eqref{eq:signal-model} by $K=1$, $f_{k,l}=\theta_k\ist l$, $\theta_1=f_0$ and $\mathcal{S}_1=\{1,\,2,\,3,\,4,\,5\}$. If several notes are played together to form a chord, the different harmonic series are superimposed on each other. Thus, the line spectrum will consist of $K>1$ such harmonic series with different fundamental pitches each.
}
\begin{align}\label{eq:signal-model}
	\bm{x} &=\sum_{k=1}^K\sum_{l \in \mathcal{S}_k} \alpha_{k,l} \, \bm{\psi}(\theta_k,l) + \bm{\epsilon} \ist .
	\\[-7mm]\nn
\end{align}
Each group consists of one or multiple spectral lines $\bm{\psi}(\theta_k,l) = e^{j2\pi f_{k,l} \bm{t}}$, also referred to as components, with frequencies $f_{k,l}$ related to the parameter $\theta_k$ by a finite discrete alphabet $l\in\mathcal{S}_k$.
\footnote{$e^{\bm{a}}$ with $\bm{a}=[a_1\iist a_2\,\cdots \, a_N]^\transpose \in \mathbb{C}^N$ is defined to be a vector, i.e., $e^{\bm{a}} \triangleq [e^{a_1}\iist e^{a_2}\,\cdots\,e^{a_N}]^\transpose$.}
Furthermore, each spectral line $\bm{\psi}(\theta_k,l)$ is weighted with an amplitude $\alpha_{k,l}\in\mathbb{C}$ and the signal is corrupted by \gls{awgn} $\bm{\epsilon}$.
We assume $\bm{\epsilon}$ to be sampled from a Gaussian random process with double sided power-spectral density $N_0/2$.
Hence, $\bm{\epsilon}$ follows a circular-symmetric complex Gaussian distribution, i.e., $p(\bm{\epsilon}) = \mathcal{CN}(\bm{\epsilon}\, |\, 0,\, \lambda^{-1}\bm{I})$ with precision $\lambda=\frac{1}{N_0}$.
\footnote{We denote the complex Gaussian \acrshort{pdf} of the variable $\bm{x}\in\mathbb{C}^N$ with mean $\bm{\mu}$ and covariance $\bm{\Sigma}$ as $\mathcal{CN}(\bm{x}|\bm{\mu},\bm{\Sigma})=|\pi \bm{\Sigma}|^{-1} \exp\{-(\bm{x}-\bm{\mu})^\hermitian\bm{\Sigma}^{-1}(\bm{x}-\bm{\mu})\}$, where $|\cdot|$ denotes the matrix determinant. Furthermore, we assume that $\bm{x}-\bm{\mu}$ is proper for all complex Gaussian random variables $\bm{x}$ with mean $\bm{\mu}$.}

We aim to estimate the number of groups $K$, the fundamental frequencies $\theta_k$, the group structure $\mathcal{S}_{k}$ of each group, the amplitudes $\alpha_{k,l}$ and noise variance $\lambda$.
Note, that the signal model in \eqref{eq:signal-model} can be straightforwardly extended to multiple measurement vectors such as signals from an microphone or antenna array and to vector parameters such as estimating the angle-of-arrival in addition to the fundamental frequency. Furthermore, any set of functions can be selected as basis instead of the structured spectral lines. Thus, the signal model is potentially applicable to an even wider variety of engineering problems.

\subsection{Inference Model and Bayesian Formulation}

To perform (approximate) Bayesian inference on this model, we rewrite \eqref{eq:signal-model} as product of a large parametrized dictionary matrix $\bm{\Psi}(\bm{\theta})$ whose columns contain all possible components of a large number of groups multiplied with a sparse amplitude vector $\bm{\alpha}$.
Let $\mathcal{S}_{\text{max}}=\{l\,|\,L_{\text{min}}\leq l \leq L_{\text{max}}\}$ be the set of all potential components of a group defined by $L_{\text{min}}$ and $L_{\text{max}}$, and $K_{\text{max}}$ the maximum number of groups.
\footnote{Since we apply a bottom-up initialization and we can never expect to estimate more parameters than the number of observations, the actual values of $L_{\text{min}}$, $L_{\text{max}}$ and $K_{\text{max}}$ do not influence the proposed algorithm as long as they allow for a large enough number of groups and components per group.}
Let $\bm{\theta}=[\theta_1 \iist \theta_2\,\cdots\,\theta_{K_{\text{max}}}]^\transpose$ be a vector of the corresponding fundamental frequencies $\theta_k$ and $\bm{\Psi}(\theta_k)=[\bm{\psi}(\theta_k,L_{\text{min}}) \iist\bm{\psi}(\theta_k,L_{\text{min}}+1)\,\cdots\,\bm{\psi}(\theta_k,L_{\text{max}})]$ a matrix whose columns contain all spectral lines parametrized by $\theta_k$, and $\bm{\Psi}(\bm{\theta})=[\bm{\Psi}(\theta_1)\iist\bm{\Psi}(\theta_1)\,\cdots\,\bm{\Psi}(\theta_{K_{\text{max}}})]$.
Furthermore, we introduce amplitude vectors for each possible group $\bm{\alpha}_{k} = [\alpha_{k,L_{\text{min}}}\iist\alpha_{k,L_{\text{min}}+1}\,\cdots\,\alpha_{k,L_{\text{max}}}]^\transpose$ and the sparse vector of all amplitudes $\bm{\alpha} = [\bm{\alpha}_1^\transpose \iist \bm{\alpha}_2^\transpose\,\cdots\,\bm{\alpha}_{K_{\text{max}}}^\transpose]^\transpose$.
With this the inference model of the signal model in \eqref{eq:signal-model} is given by
\begin{align}
	\bm{x}
	&=\bm{\Psi}(\bm{\theta})\bm{\alpha} + \bm{\epsilon}\ist.
	\label{eq:signal-model-inference}\\[-7mm]\nn
\end{align}

\begin{figure}
	\centering
	\includegraphics{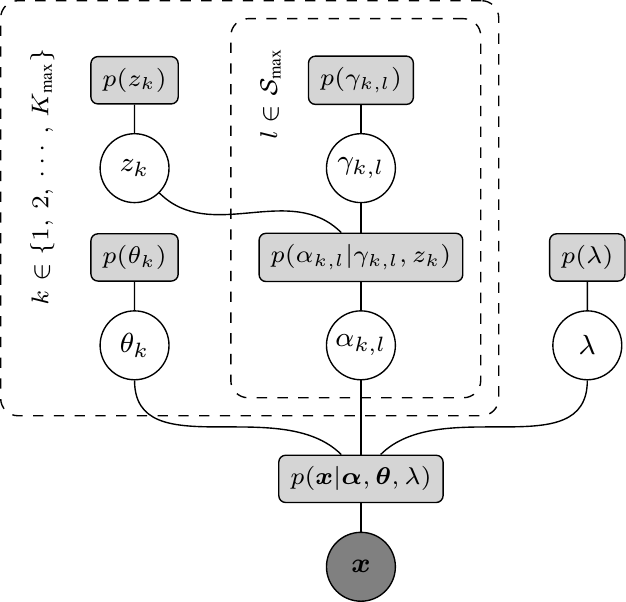}
	\caption{Factor graph representation of the Bernoulli-Gamma-Gaussian model for sparse group estimates. Sparsity of components is promoted by Gamma \glspl{pdf} as hyperpriors for $\gamma_{k,l}$, while sparsity in groups is promoted by Bernoulli \glspl{pdf} as hyperpriors for $z_k$.}
	\label{fig:graphical_model_simple}
\end{figure}

To achieve sparsity on both levels, in the number of groups as well as in the number of components in each group, we propose to use a Bernoulli-Gamma-Gaussian prior model.
We model the existence of each group with independent Bernoulli distributed random variables while simultaneously modeling the prior variance of each amplitude with Gamma distributed random variables.
A factor graph representation of the model is depicted in Figure \ref{fig:graphical_model_simple}.
Our model differs from \cite{shutin2013:VSBL,hansen2014SAM:SBL} in the addition of the Bernoulli-prior which is shown to increase resilience against the insertion of artificial components \cite{badiu2017TSP:VSBL}, while it differs from \cite{hansen2018TSP:SuperFastLSE,badiu2017TSP:VSBL} by using the Bernoulli-prior to model the existence of groups of several components instead of individual components.
Note, that we can constrain each group to contain at most a single spectral line with frequency $\theta_k$ by setting $\mathcal{S}_{\text{max}} = \{0\}$ and $f_{k,l}=\theta_k$.
In this case, the hierarchical model is identical to \cite{hansen2018TSP:SuperFastLSE}.
Therefore, the presented method can be viewed as a generalization of \cite{hansen2018TSP:SuperFastLSE}, except we use a variational-\gls{em} inference scheme instead of maximizing a Type-II likelihood function \cite{wipf2011TIP}.
We would like to emphasise here that the hierarchical model is just a ``convenient fiction'' in order to construct useful cost functions for penalized regression of the form
\begin{align}
	\hat{\bm{\theta}},\hat{\bm{\alpha}}=\arg\min_{\bm{\theta},\bm{\alpha}} \|\bm{x}-\bm{\Psi}(\bm{\theta})\bm{\alpha}\|^2 + g(\bm{\theta},\bm{\alpha})\\[-7mm]\nn
\end{align}
where $g(\bm{\theta},\bm{\alpha})$ is a penalty term which promotes sparsity \cite{wipf2011TIP}.

For each potential group $k\in \{1,\,2,\,\cdots,\,K_{\text{max}}\}$, we introduce binary random variables $\bm{z}=[z_1 \iist z_2\,\cdots\,z_{K_{\text{max}}}]^\transpose$, $z_k\in\{0,\,1\}$ which select whether the $k$-th group is active or not. If the $k$-th group is not active all amplitudes are zero $\alpha_{k,l}=0\, \forall\, l$.
The prior \gls{pdf} of the amplitudes $\alpha_{k,l}$ of all active groups is further modeled by independent complex Gaussian \glspl{pdf} with precisions $\gamma_{k,l}$, that are again treated as random variables and inferred as nuisance parameters \cite{hansen2014SAM:SBL,shutin2013:VSBL, hansen2018TSP:SuperFastLSE,tipping1999NeurIPS:RelevanceVector,tipping2003WAIS:FastMarginalSparseBayesian}.
Thus, the prior \gls{pdf} for an individual amplitude $\alpha_{k,l}$ conditioned on $z_k$ and $\gamma_{k,l}$ is then given by
\vspace*{-0.5mm}
\begin{align}
	\hspace*{-1mm} p(\alpha_{k,l}|\gamma_{k,l},\,z_k) =
	z_k \ist \mathcal{CN}(\alpha_{k,l}|0,\,\gamma_{k,l}^{-1}) + (1-z_k)\ist\delta(\alpha_{k,l})
	\label{eq:ampltiude-prior}\\[-5mm]\nn
\end{align}
where $\delta(\cdot)$ is the Dirac delta function.
The variables $z_k$ represent sparsity-inducing priors on the group level and their hyperpiors are modeled as independent Bernoulli \glspl{pdf} with $\bar{z}$ as probability for $z_k=1$, i.e.,
\vspace*{-1mm}
\begin{align}\label{eq:BernoulliPrior}
	p(\bm{z}) = \prod_{k=1}^{K_{\text{max}}}\bar{z}^{z_k} \ist (1-\bar{z})^{1-z_k}\ist.\\[-7mm]\nn
\end{align}
The precisions $\gamma_{k,l}$ represent the prior variances for components within a group and their hyperpriors are modeled by independent Gamma \glspl{pdf} $p(\gamma_{k,l})=\text{Ga}(\gamma_{k,l}|\eta,\nu)$ with shape $\eta$ and rate $\nu$.
\footnote{We denote the Gamma \acrshort{pdf} with shape $a$ and rate $b$ as $\text{Ga}(x\,|\, a,b)=\frac{b^a}{\Gamma(a)}x^{a-1}e^{-b x}$, where $\Gamma(\cdot)$ is the gamma function.}
Note that since the \glspl{pdf} of the amplitude's precisions $p(\gamma_{k,l})$ are sparsity inducing, the according hierarchical model leads to many component amplitudes having a prior variance of zero, resulting in them being removed from the model as the corresponding amplitude is estimated to be zero as well. 
The number of groups and number of components within a group are indirectly estimated by, respectively, estimating the posterior \glspl{pdf} of $z_k$ and $\gamma_{k,l}$.

Let  $\bm{\gamma}_k=[\gamma_{k,L_{\text{min}}} \iist\gamma_{k,L_{\text{min}}+1}\,\cdots\,\gamma_{k,L_{\text{max}}}]^\transpose$ be a column-vector of the precisions $\gamma_{k,l}$ corresponding to the amplitudes $\bm{\alpha}_{k}$ of the $k$-th group and $\bm{\gamma}=[\bm{\gamma}_1^\transpose\iist\bm{\gamma}_2^\transpose\,\cdots\,\bm{\gamma}_{K_{\text{max}}}^\transpose]^\transpose$ be a vector of all precisions $\gamma_{k,l}$.
Let $\mathcal{S}=\mathcal{S}(\bm{\gamma},\bm{z})$ be an index set such that $\bm{\alpha}_{\mathcal{S}}$ contains all nonzero elements of $\bm{\alpha}$ and $\bm{\gamma}_{\mathcal{S}}$ contains the prior variances corresponding the amplitudes $\bm{\alpha}_{\mathcal{S}}$.
\footnote{We denote a vector subscripted by an index set $\bm{\alpha}_{\mathcal{S}}$ as the vector containing the elements of $\bm{\alpha}$ whose indices are elements of $\mathcal{S}$. Similarly, we denote for matrices $\bm{\Psi}_{\mathcal{S}}(\bm{\theta})$ as the matrix formed by the columns of $\bm{\Psi}(\bm{\theta})$ whose indices are elements of $\mathcal{S}$.}
Finally, let $\bm{\Gamma}=\text{diag}(\bm{\gamma}_{\mathcal{S}})$, where $\text{diag}(\bm{\gamma}_{\mathcal{S}})$ denotes a diagonal matrix with the elements of the vector $\bm{\gamma}_{\mathcal{S}}$ along the main diagonal, such that the joint prior \gls{pdf} of the amplitudes is given by
\vspace*{-0.5mm}
\begin{align}
	p(\bm{\alpha}|\bm{\gamma},\bm{z}) 
	&=\mathcal{CN}(\bm{\alpha}_{\mathcal{S}}|\bm{0},\,\bm{\Gamma}^{-1}) \prod_{\alpha_{k,l}\notin \bm{\alpha}_{\mathcal{S}}} \delta(\alpha_{k,l})\ist.
	\label{eq:prior-alpha}\\[-6mm]\nn
\end{align}
We assume the prior distribution of the noise variance to be a Gamma \gls{pdf} $p(\lambda)=\text{Ga}(\lambda\,|\,\rho,\,\mu)$  with shape $\rho$ and rate $\mu$, since this is the conjugate prior for the variance of a Gaussian \gls{pdf}.
From the \gls{awgn} assumption it follows that the likelihood $p(\bm{x}|\bm{\alpha},\bm{\theta},\lambda)$ follows a Gaussian \gls{pdf}
\vspace*{-0.5mm}
\begin{align}
	p(\bm{x}|\bm{\alpha},\bm{\theta},\lambda) = \mathcal{CN}(\bm{x}\, |\, \bm{\Psi}_{\mathcal{S}}(\bm{\theta})\bm{\alpha}_{\mathcal{S}},\, \lambda^{-1}\bm{I})\ist.
	\label{eq:likelihood}\\[-6mm]\nn
\end{align}
Introducing $p(\bm{\theta})$ as the prior for the parameters $\bm{\theta}$ and using the Bayes theorem, the posterior \gls{pdf} is proportional to
\vspace*{-0.5mm}
\begin{align}
	p(\bm{\alpha},\bm{\theta},\bm{\gamma},&\bm{z},\lambda|\bm{x})  \nonumber\\
	&\propto p(\bm{x}|\bm{\alpha},\bm{\theta},\lambda) p(\bm{\alpha}|\bm{\gamma},\bm{z}) p(\bm{\gamma}) p(\bm{z}) p(\bm{\theta}) p(\lambda)\ist.
	\label{eq:posterior-distribution}\\[-6mm]\nn
\end{align}
Calculating a maximum a-posteriori estimate from \eqref{eq:posterior-distribution} is computationally prohibitive for all but the most simplest problems of interest due to the high dimensionality, interdependencies between variables and nonlinearities in the model. Thus, we apply a variational-\gls{em} approach \cite{tzikas2008:VAEM}, \cite[Ch. 10]{Bishop2006} together with a structured mean-field assumption to approximate the posterior \gls{pdf}.

\section{Variational Approximation}
\label{sec:variational-solution}

\subsection{Mean-Field Factorization and Distribution Updates}

We consider $\bm{\theta}$ and $\bm{z}$ as deterministic unknowns and estimate point estimates $\hat{\bm{\theta}}$ and $\hat{\bm{z}}$, while we approximate the posterior distribution with a factorized proxy \gls{pdf} given by
\begin{align}
	q(\bm{\alpha},\bm{\gamma},\lambda;\bm{\theta},\bm{z})&=q_{\bm{\alpha}}(\bm{\alpha};\bm{\theta},\bm{z})\,q_{\lambda}(\lambda;\bm{\theta},\bm{z}) \nonumber \\
	&\quad \times \prod_{k=1}^{K_{\text{max}}}\prod_{l=L_{\text{min}}}^{L_{\text{max}}}q_{\gamma,k,l}(\gamma_{k,l};\bm{\theta},\bm{z})	
\end{align}
parametrized by $\bm{\theta}$ and $\bm{z}$. This factorized \gls{pdf} consists of a joint proxy \gls{pdf} $q_{\bm{\alpha}}$ for all amplitudes and independent proxy \glspl{pdf} $q_{\gamma,k,l}$ and $q_{\lambda}$ for all prior variances $\gamma_{k,l}$ and the noise precision $\lambda$, respectively.
We do not constrain the factors of the proxy \gls{pdf} to be from a specific family.
Thus, their shape is determined by the variational optimization procedure.
We minimize the \gls{kl}-divergence between the true posterior \gls{pdf} $p(\bm{\alpha},\bm{\theta},\bm{\gamma},\bm{z},\lambda|\bm{x})$ and the proxy \gls{pdf} $q(\bm{\alpha},\bm{\gamma},\lambda;\bm{\theta},\bm{z})$ by maximizing the \gls{elbo} \cite{tzikas2008:VAEM}, \cite[Ch. 10]{Bishop2006}
\begin{align}
	\mathcal{L}(q,\bm{\theta},&\bm{z})\nonumber \\
	&  = \big<\ln p(\bm{\alpha},\bm{\theta},\bm{\gamma},\bm{z},\lambda|\bm{x})-\ln q(\bm{\alpha},\bm{\gamma},\lambda;\bm{\theta},\bm{z})\big>_q
	\label{eq:elbo}
\end{align}
where $\big<f(x)\big>_{q(x)}$ denotes the expectations of the function $f(x)$ with respect to the random variable $x$ distributed according to the proxy \gls{pdf} $q(x)$.
We alternate between M-steps to maximize \gls{elbo} with respect to one or several of the proxy \glspl{pdf} $q_j \in \mathcal{Q} = \{q_{\bm{\alpha}},\,q_{\lambda},\,q_{\gamma,1,L_{\text{min}}},\,q_{\gamma,1,L_{\text{min}}+1},\,\cdots,\,q_{\gamma,K_{\text{max}},L_{\text{max}}}\}$, and E-Steps to maximize the \gls{elbo} with respect to the parameters $\hat{\bm{\theta}}$ and $\hat{\bm{z}}$, based on the updated proxy \glspl{pdf} from the M-step.
To avoid cluttered notation, we omit explicit iteration indices and refer to the last available estimates of the respective proxy \glspl{pdf} and parameters.

In the M-step, each proxy \gls{pdf} $q_j\in \mathcal{Q}$ is calculated according to
\begin{equation}
	q_j \propto \exp \big<\ln p(\bm{\alpha},\bm{\theta}=\hat{\bm{\theta}},\bm{\gamma},\bm{z}=\hat{\bm{z}},\lambda|\bm{x})\big>_{\bar{q}_{j}}
	\label{eq:message_update}
\end{equation}
where $\bar{q}_{j}=\prod_{q_i \in \mathcal{Q}\backslash q_j} q_i\vspace*{0.5mm}$ denotes the product of all factors of the joint proxy \gls{pdf} $q$ except $q_j$. Let $\hat{\bm{\gamma}}$ be the estimated mean value of $\bm{\gamma}$ based on all $q_{\gamma,k,l}$ and $\hat{\mathcal{S}}=\mathcal{S}(\hat{\bm{\gamma}},\hat{\bm{z}})$ the current estimate of $\mathcal{S}$.
As we derive in Appendix \ref{sec:appendix:update-equations}, inserting \eqref{eq:posterior-distribution} into \eqref{eq:message_update} results in the proxy \glspl{pdf}
\begin{align}
	q_{\bm{\alpha}}(\bm{\alpha};\hat{\bm{\theta}},\hat{\bm{z}}) &= \mathcal{CN}(\bm{\alpha}_{\hat{\mathcal{S}}}\, |\, \hat{\bm{\alpha}},\, \hat{\bm{C}}) \prod_{\alpha_{k,l}\notin \bm{\alpha}_{\hat{\mathcal{S}}}} \delta(\alpha_{k,l})
	\label{eq:q-alpha} \\
	q_\lambda(\lambda;\hat{\bm{\theta}},\hat{\bm{z}}) &= \text{Ga}(\lambda\, |\, N+\rho,\, \hat{M}_\lambda) 
	\label{eq:q-lambda} \\
	q_{\gamma,k,l}(\gamma_{k,l};\hat{\bm{\theta}},\hat{\bm{z}}) &= \begin{cases}
		\text{Ga}(\gamma_{k,l}\, |\, \eta+1,\, \hat{M}_{k,l}) & \text{if } z_k=1 \\
		p(\gamma_{k,l}) & \text{if } z_k=0
	\end{cases}
	\label{eq:q-gamma}
\end{align}
where $\hat{M}_{\lambda}$ and $\hat{M}_{k,l}$ are the rate parameters of the resulting Gamma \glspl{pdf} for $q_{\lambda}$ and $q_{\gamma,k,l}$.
If $\hat{z}_k=0$, then $q_{\gamma,k,l}(\gamma_{k,l};\hat{\bm{\theta}},\hat{\bm{z}})=p(\gamma_{k,l})$, which means that the whole group is deactivated based the Bernoulli-prior model.
Let $\text{tr}(\cdot)$ denote the matrix trace operator, $\hat{\bm{\Psi}}=\bm{\Psi}_{\hat{\mathcal{S}}}(\hat{\bm{\theta}})$ be the matrix of all spectral lines with nonzero amplitudes parametrized by $\hat{\bm{\theta}}$, $\hat{\bm{\Gamma}}=\text{diag}(\hat{\bm{\gamma}}_{\hat{\mathcal{S}}})$ a diagonal matrix with the priors $\hat{\gamma}_{k,l}$ of the active components on its main diagonal and $\hat{C}_{k,l}$ the element on the main diagonal of $\hat{\bm{C}}$ that corresponds to the estimated variance of the amplitude $\alpha_{k,l}$. 
Thus, the parameters of the proxy \glspl{pdf} in \eqref{eq:q-alpha}, \eqref{eq:q-lambda}, and \eqref{eq:q-gamma} are given, respectively, by
\begin{align}
	\hat{\bm{\alpha}} &= \hat{\lambda}\hat{\bm{C}}\hat{\bm{\Psi}}^\hermitian\bm{x}\, , \label{eq:update_alpha} \\
	\hat{\bm{C}} &= \big(\hat{\lambda}\hat{\bm{\Psi}}^\hermitian\hat{\bm{\Psi}} + \hat{\bm{\Gamma}}\big)^{-1}\, , \label{eq:update_alphaCovar}
\end{align}
\begin{align}
	\hat{\lambda} &= \frac{N+\rho}{\|\bm{x}-\hat{\bm{\Psi}}\hat{\bm{\alpha}}\|^2 + \text{tr}(\hat{\bm{\Psi}}\hat{\bm{C}}\hat{\bm{\Psi}}^\hermitian)+\mu}\, ,
	\label{eq:update_lambda}
\end{align}
and
\begin{align}
	\hat{\gamma}_{k,l} &= \begin{cases}
		\frac{\eta + 1}{\hat{C}_{k,l} + \|\hat{\alpha}_{k,l}\|^2 + \nu} & \text{if } \hat{z}_k=1 \\
		\text{not defined} & \text{if } \hat{z}_k=0
	\end{cases}
	\label{eq:update_gamma_single}.
\end{align}

\subsection{Fundamental Frequencies $\theta$ and Group Activations $z$}
\label{sec:variational-solution:group-param-update}

In order to estimate the fundamental frequencies $\hat{\bm{\theta}}$ and active groups $\hat{\bm{z}}$, the \gls{elbo} is maximized jointly with respect to $q_{\bm{\alpha}}$, $\bm{\theta}$, and $\bm{z}$ \cite{badiu2017TSP:VSBL}.
In the following we use the product of all $q_{\gamma,k,l}$, i.e., $q_{\bm{\gamma}}=\prod_{k=1}^{K_{\text{max}}} \prod_{l=L_{\text{min}}}^{L_{\text{max}}}q_{\gamma,k,l}$ and $\overset{e}{\propto}$, which means that the right side is equal to the left side plus a constant, such that both sides are proportional to each other after applying the exponential function.
The \gls{elbo} in \eqref{eq:elbo} for the proxy \gls{pdf} $q_\alpha$ is given by
\begin{align}
	\mathcal{L}(&q_\alpha, \bm{\theta}, \bm{z}) \nonumber\\
	&\overset{e}{\propto} \big<\big<\ln p(\bm{\alpha},\bm{\theta},\bm{\gamma},\bm{z},\lambda|\bm{x})\big>_{q_{\lambda}q_{\bm{\gamma}}}
	 \hspace{-0.5em} - \ln q_{\bm{\alpha}}(\bm{\alpha};\bm{\theta},\bm{z})\big>_{q_{\bm{\alpha}}}
	\label{eq:elbo_marginal}
	\,.\\[-7mm]\nonumber
\end{align}
Since ${\bm{\theta}}$ and ${\bm{z}}$ are restricted to point estimates, the standard free-form optimization \cite[Ch. 10]{Bishop2006} is not applicable. Following \cite{badiu2017TSP:VSBL}, we introduce the \gls{pdf}
\begin{equation}
	t(\bm{\alpha};\bm{\theta},\bm{z}) = \frac{1}{Z(\bm{\theta},\bm{z})} \exp \big<\ln p(\bm{\alpha},\bm{\theta},\bm{\gamma},\bm{z},\lambda|\bm{x})\big>_{q_{\lambda}q_{\bm{\gamma}}}
	\label{eq:helperPDF}
\end{equation}
with normalization constant
\begin{equation}
	Z(\bm{\theta},\bm{z}) = \int_{\bm{\alpha}}\exp\big<\ln p(\bm{\alpha},\bm{\theta},\bm{\gamma},\bm{z},\lambda|\bm{x})\big>_{q_{\lambda}q_{\bm{\gamma}}} d\bm{\alpha}
	\label{eq:activations_distribution_t}\,.
\end{equation}
Using \eqref{eq:helperPDF} and \eqref{eq:activations_distribution_t}, the \gls{elbo} in \eqref{eq:elbo_marginal} can be rewritten as
\begin{equation}
	\mathcal{L}(q_{\bm{\alpha}},\bm{\theta},\bm{z}) = \text{const.}-\mathcal{D}_{\text{KL}}(q_{\bm{\alpha}}\|t) + \ln Z(\bm{\theta},\bm{z})
	\label{eq:elbo_KL-div_theta_z}
\end{equation}
where $\mathcal{D}_{\text{KL}}(q\|p)$ denotes the \gls{kl} divergence of $q$ from $p$. Since $\mathcal{D}_{\text{KL}}\geq 0$ with equality if and only if $t=q_{\bm{\alpha}}$, \eqref{eq:elbo_KL-div_theta_z} is maximized by
\vspace*{-1mm}
\begin{align}
	q_{\bm{\alpha}}(\bm{\alpha};\hat{\bm{\theta}},\hat{\bm{z}}) = t(\bm{\alpha};\hat{\bm{\theta}},\hat{\bm{z}})\\[-7mm]\nonumber
\end{align}
where point estimates $\hat{\bm{\theta}}$ and $\hat{\bm{z}}$ are determined by
\begin{align}
	\hat{\bm{\theta}},\hat{\bm{z}}=\arg\max_{\bm{\theta},\bm{z}} \ln Z(\bm{\theta},\bm{z})\,.
	\label{eq:elbo_maxmization_ansatz}
\end{align}
Let $\tilde{\mathcal{S}}=\mathcal{S}(\hat{\bm{\gamma}},\bm{z})$, $\bm{\Psi}_{\tilde{\mathcal{S}}}=\bm{\Psi}_{\tilde{\mathcal{S}}}(\bm{\theta})$, $\hat{\bm{\Gamma}}_{\tilde{\mathcal{S}}}=\text{diag}(\hat{\bm{\gamma}}_{\tilde{\mathcal{S}}})$ and $\bm{C}_{\tilde{\mathcal{S}}}=(\hat{\lambda}\bm{\Psi}_{\tilde{\mathcal{S}}}^\hermitian\bm{\Psi}_{\tilde{\mathcal{S}}} + \hat{\bm{\Gamma}}_{\tilde{\mathcal{S}}})^{-1}$.
As we derive in Appendix \ref{sec:appendix-lnZ-maximization}, we find $\hat{\bm{\theta}}$ and $\hat{\bm{z}}$ from \eqref{eq:elbo_maxmization_ansatz} as the maximizer of
\begin{align}
	\ln Z(\bm{\theta},\bm{z}) &\overset{e}{\propto} \hat{\lambda}^2\bm{x}^\hermitian \bm{\Psi}_{\tilde{\mathcal{S}}}\bm{C}_{\tilde{\mathcal{S}}}\bm{\Psi}_{\tilde{\mathcal{S}}}^\hermitian\bm{x} + \ln | \bm{C}_{\tilde{\mathcal{S}}}| + \ln p(\bm{z}) \nonumber \\
	&\qquad + \sum_{\hat{\gamma}_{k,l} \in \hat{\bm{\gamma}}_{\tilde{\mathcal{S}}}} \big<\ln \gamma_{k,l}\big>_{q_{\gamma,k,l}}  + \ln p(\bm{\theta})
	\label{eq:elbo_maximization_global}\\[-7mm]\nonumber
\end{align}
which is similar to the Type-II cost function that is obtained by maximizing the marginalized likelihood \cite{hansen2014SAM:SBL}.
Finding the global maximum over all possible values of $\bm{\theta}$ and $\bm{z}$ is computationally prohibitive.
Therefore, we express the dependence on one set of parameters $z_{k}$, $\theta_k$ explicitly and maximize \eqref{eq:elbo_maximization_global} by coordinate ascent.

Let $\bm{\Psi}_k$ denote the columns of $\bm{\Psi}_{\tilde{\mathcal{S}}}$ which correspond to the $k$-th group and let the index $\bar{k}$ refer to all the columns of matrices, or elements of vectors, which do not correspond to the $k$-th group.
Without loss of generality, we can reorder $\hat{\bm{\gamma}}_{\tilde{\mathcal{S}}} = [\hat{\bm{\gamma}}_{\bar{k}}^\transpose\iist\ist\hat{\bm{\gamma}}_{k}^\transpose]^\transpose$, $\bm{\theta} = [\bm{\theta}_{\bar{k}}^\transpose\iist\ist\theta_k]^\transpose$, $\bm{z} = [\bm{z}_{\bar{k}}^\transpose\iist\ist z_k]^\transpose$, and $\bm{\Psi}_{\tilde{\mathcal{S}}} = [\bm{\Psi}_{\bar{k}}\iist\ist\bm{\Psi}_k]$ such that the elements corresponding to the $k$-th group are moved to the end.
Let $\hat{\bm{\Gamma}}_{\bar{k}}=\text{diag}(\hat{\bm{\gamma}}_{\bar{k}})$, and $\hat{\bm{\Gamma}}_k=\text{diag}(\hat{\bm{\gamma}}_k)$, $\bm{C}_{\bar{k}} = (\hat{\lambda}\bm{\Psi}_{\bar{k}}^\hermitian\bm{\Psi}_{\bar{k}}+\hat{\bm{\Gamma}}_{\bar{k}})^{-1}$, $\bm{C}_k = (\hat{\lambda}\bm{\Psi}_k^\hermitian\bm{\Psi}_k + \hat{\bm{\Gamma}}_k - \hat{\lambda}^2 \bm{\Psi}_k^\hermitian\bm{\Psi}_{\bar{k}}\bm{C}_{\bar{k}}\bm{\Psi}_{\bar{k}}^\hermitian\bm{\Psi}_k)^{-1}$, and $\bm{u} =\hat{\lambda}\bm{C}_k\bm{\Psi}_k^\hermitian(\bm{I}-\hat{\lambda}\bm{\Psi}_{\bar{k}}\bm{C}_{\bar{k}}\bm{\Psi}_{\bar{k}}^\hermitian)\bm{x}$.
As detailed in Appendix \ref{sec:appendix:update-equations}, we assume the parameter priors $p(\bm{\theta})$ to be independent and flat, to express the difference between $\ln Z$ with the $k$-th group removed and $\ln Z$ including the $k$-th group as
\begin{align}
	&\Delta_k(\theta_k) \nn \\
	&=\ln Z\big([\hat{\bm{\theta}}_{\bar{k}}^\transpose\iist\ist\theta_k]^\transpose,[\hat{\bm{z}}_{\bar{k}}^\transpose\iist\ist 1]^\transpose\big) - \ln Z\big([\hat{\bm{\theta}}_{\bar{k}}^\transpose\iist\ist\theta_k]^\transpose,[\hat{\bm{z}}_{\bar{k}}^\transpose\iist\ist 0]^\transpose\big)
	\nonumber \\
	&=  \bm{u}^\hermitian\bm{C}_k^{-1}\bm{u} + \ln |\bm{C}_k| + \ln\frac{\bar{z}}{1-\bar{z}} +\rmv \sum_{l \in \hat{\mathcal{S}}_k}\rmv\big(\chi_0\rmv+\rmv\ln \hat{\gamma}_{k,l}\big)
	\label{eq:elbo_difference_group}\\[-7mm]\nonumber
\end{align}
where $\hat{\mathcal{S}}_k= \{l\ist |\ist \hat{\gamma}_{k,l} < \infty\}$ and $\chi_0=\text{digamma}(\eta+1)$ is the digamma function evaluated at $\eta+1$.
After finding $\hat{\theta}_{k}=\arg\max_{\theta_k}\Delta_k(\theta_k)$, we activate the $k$-th group by $\hat{z}_k=1$ and update the respective parameter to $\hat{\theta}_{k}$ if $\Delta_k(\hat{\theta}_{k})>0$, indicating an increase in $\ln Z$ compared to deactivating the $k$-th group by $\hat{z}_k=0$.

\subsection{Fast Update of Priors $\gamma_{k,l}$ and Component Threshold $\chi_1$}
\label{sec:variational-solution:fast-gamma-update}

If the prior \glspl{pdf} $p(\gamma_{k,l})$ is sparsity-inducing, many estimates $\hat{\gamma}_{k,l}$ will diverge if the updated equations \eqref{eq:update_alpha} trough \eqref{eq:update_gamma_single} are iterated ad infinitum, resulting in a sparse estimate for $\hat{\bm{\alpha}}$ \cite{tipping1999NeurIPS:RelevanceVector,shutin2011TSP:fastVSBL}.
To obtain a fast convergence check, we consider Jeffery's prior $p(\gamma_{k,l})\propto \gamma_{k,l}^{-1}$ obtained by $\eta=\nu=0$ and investigating the dependency of $\hat{\gamma}_{k,l}$ on $\hat{\bm{\alpha}}$ and $\hat{\bm{C}}$.
Following \cite{shutin2011TSP:fastVSBL}, we can express repeated cycles of updating $q_{\bm{\alpha}}$ followed by updating $q_{\gamma,k,l}$ as a nonlinear map. Inserting \eqref{eq:update_alpha} and \eqref{eq:update_alphaCovar} into \eqref{eq:update_gamma_single}, each cycle $i$ maps from the previous estimate of $\hat{\gamma}_{k,l}$ to the next as $\hat{\gamma}_{k,l}^{[i]} = F(\hat{\gamma}_{k,l}^{[i-1]})$.
Hence, we can derive fast update rules for $\gamma_{k,l}$ by analysing the stationary points of the map $F(\cdot)$. 
Let $\bm{\psi}_{k,l}=\bm{\psi}(\hat{\theta}_k,l)$, $\bm{\Psi}_{\overline{k,l}}$ be the dictionary matrix $\hat{\bm{\Psi}}$ with the column $\bm{\psi}_{k,l}$ removed, $\hat{\bm{\Gamma}}_{\overline{k,l}}$ a diagonal matrix containing the elements of $\bm{\hat{\gamma}}_{\hat{\mathcal{S}}}$ with $\hat{\gamma}_{k,l}$ removed and $\bm{C}_{\overline{k,l}}=\big(\hat{\lambda}\bm{\Psi}_{\overline{k,l}}^\hermitian\bm{\Psi}_{\overline{k,l}} + \hat{\bm{\Gamma}}_{\overline{k,l}}\big)^{-1}$.
Furthermore, let
$s_{k,l} = (\hat{\lambda}\bm{\psi}_{k,l}^\hermitian \bm{\psi}_{k,l} - \hat{\lambda}^2 \bm{\psi}_{k,l}^\hermitian \bm{\Psi}_{\overline{k,l}}\bm{C}_{\overline{k,l}}\bm{\Psi}_{\overline{k,l}}^\hermitian \bm{\psi}_{k,l})^{-1}$ and $u_{k,l} = s_{k,l} (\hat{\lambda} \bm{\psi}_{k,l}^\hermitian\bm{x} - \hat{\lambda}^2 \bm{\psi}_{k,l}^\hermitian \bm{\Psi}_{\overline{k,l}}\bm{C}_{\overline{k,l}}\bm{\Psi}_{\overline{k,l}}^\hermitian\bm{x})$, the map $F(\cdot)$ can be shown to converge to
\vspace*{-0.5mm}
\begin{align}
	\hat{\gamma}_{k,l} \triangleq \hat{\gamma}_{k,l}^{[\infty]} = (|u_{k,l}|^2-s_{k,l})^{-1} \quad \text{if} \quad
	\frac{|u_{k,l}|^2}{s_{k,l}} > 1
	\label{eq:gamma_convergence_single}\\[-7mm]\nonumber
\end{align}
and diverges otherwise.
Thus, if \eqref{eq:gamma_convergence_single} is fulfilled we keep the $l$-th component of the $k$-th group in the model and discard it otherwise.
A similar analysis can be performed for $\eta > 0$ and $\nu > 0$. However, for the sake of brevity we consider this analysis to be outside the scope of this work.

It can be shown that $\frac{|u_{k,l}|^2}{s_{k,l}}$ corresponds to the component \gls{snr} \cite{shutin2011TSP:fastVSBL} and, thus, the condition $\frac{|u_{k,l}|^2}{s_{k,l}} > 1$ equals accepting any component that is even slightly above the noise level. However, this will also result in some false alarms.
We can heuristically increase the threshold to $\frac{|u_{k,l}|^2}{s_{k,l}} > \chi_1 \geq  1$ in order to reduce the false alarm rate at the cost of an increased missed detection rate, where $\chi_1$ corresponds to the minimum required component \gls{snr}.
We refer the reader to \cite{leitinger2020Asilomar} for a closer analysis of the relationship between the false alarm rate and the threshold in the case of unstructured line spectra.
Note, that by increasing the threshold we lose the guarantee that each update step increases the \gls{elbo} and, thus, the guarantee for convergence.
Nevertheless, increasing the threshold was not observed to impact the performance or convergence behaviour in a noticeable manner in our simulations.

\subsection{Model Ambiguity and Constraints on Sparsity Parameters}

The model \eqref{eq:signal-model-inference} is ambiguous since many combinations of groups and active components within each group can lead to the same spectral lines. For example, if the components in each group are spectral lines with frequency $ \theta_k\ist l$, then each group can also be parametrized by $\theta_k^\prime=\frac{\theta_k}{2}$ and $l^\prime=2l$.
This effect is also known as the halfling problem \cite{swaerd2018TASLP}. We try to reduced this type of error by using a bottom-up initialization strategy as described in Section \ref{sec:algorithm}.

Furthermore, if several components are assigned to one group, we can always remove one component to form a new group, parametrized such that it results in the same spectral lines as before, increasing the degrees of freedom in the model.
Intuitively, we need to be stricter in adding new groups to the model compared to adding components within a group to avoid over parametrization.
Following \cite{hansen2018TSP:SuperFastLSE}, we express
$\bm{C}_k=\big[(s_{k,l}^{-1}+\hat{\gamma}_{k,l})^{-1}\big]=\big[\frac{s_{k,l}}{1+\hat{\gamma}_{k,l}\,s_{k,l}}\big]$ and $\bm{u} = \big[\frac{u_{k,l}}{1+\hat{\gamma}_{k,l}\,s_{k,l}}\big]\vspace*{0.5mm}$ for a group containing only a single spectral line.
From \eqref{eq:elbo_difference_group} it follows, that we activate such a group if
\begin{align}
	\frac{|u_{k,l}|^2}{s_{k,l}} \frac{1}{1+\hat{\gamma}_{k,l}\,s_{k,l}} +  \ln\frac{\hat{\gamma}_{k,l}\,s_{k,l}}{1+\hat{\gamma}_{k,l}\,s_{k,l}}+\ln\frac{\bar{z}}{1-\bar{z}} + \chi_0 > 0
	\label{eq:group-acceptance-1}
\end{align}
which depends not only on the component \gls{snr} $\frac{|u_{k,l}|^2}{s_{k,l}}$ but also on the prior $\hat{\gamma}_{k,l}$ and variance $s_{k,l}$. 
Comparing \eqref{eq:group-acceptance-1} to \eqref{eq:gamma_convergence_single}, we ensure that the inclusion of new groups with only a single component is penalized more than the inclusion of new components within a group by choosing the group existence prior $\bar{z}$ such that
\begin{equation}
	\big(1+\hat{\gamma}_{k,l}s_{k,l}\big)\big(\ln\frac{1+\hat{\gamma}_{k,l}\,s_{k,l}}{\hat{\gamma}_{k,l}\,s_{k,l}}+\ln\frac{1-\bar{z}}{\bar{z}} - \chi_0\big) > \chi_1
	\label{eq:group-acceptance-2}
\end{equation}
holds for any value of $\hat{\gamma}_{k,l}\,s_{k,l}$.
Since $\hat{\gamma}_{k,l}$ and $s_{k,l}$ are both strictly positive quantities we have $(1+\hat{\gamma}_{k,l}s_{k,l}) > 1$ and $\ln\frac{1+\hat{\gamma}_{k,l}\,s_{k,l}}{\hat{\gamma}_{k,l}\,s_{k,l}} > 0$.
Thus, \eqref{eq:group-acceptance-2} holds for any value of
\begin{equation}
	\bar{z} < \frac{1}{1+\exp(\chi_0 + \chi_1)}
	\label{eq:threshold-relation}
	.
\end{equation}
Finally, we express the cluster existence prior in terms of a second threshold $\chi_2$ as $\bar{z}=\frac{1}{1+\exp(\chi_0 + \chi_2)}$, which has to satisfy $\chi_2 > \chi_1$, for easier interpretation.

\section{Algorithm Implementation}
\label{sec:algorithm}

Updating the parameters $\hat{\bm{\theta}}$ and $\hat{\bm{z}}$ as well as the proxy distributions $q_j \in \mathcal{Q}$ in the way described in the previous section will converge towards a local optimum of the \gls{elbo}.
However, there might exist several local optima and the obtained solution depends on the initialization as well as the order in which the updates are performed.
In this section, we define an iterative schedule for updating the factors $q_{\bm{\alpha}}$, $q_{\lambda}$ and $q_{\gamma,k,l}$ and to estimate $\hat{\bm{\theta}}$ and $\hat{\bm{z}}$ as well as an initialization.
The resulting algorithm is outlined in Algorithm \ref{alg:sbcl}.
We choose Jeffrey's priors ($\rho=\mu=\eta=\nu=0$) for $p(\gamma_{k,l})$ and $p(\lambda)$, since these priors are non informative for the noise precision and it allows us to use the fast convergence check developed in Section \ref{sec:variational-solution:fast-gamma-update} for the variances $\hat{\gamma}_{k,l}$.

Without loss of generality, we can reorder the groups such that
$\hat{\bm{z}}=[1\,\cdots\,1\iist0\,\cdots\,0]$ is a vector of $\hat{K}$ leading ones followed by $K_{\text{max}}-\hat{K}$ zeros.
Therefore, we only need to keep track of the estimated number of active groups $\hat{K}$ and their parameters $\hat{\theta}_{1}$ trough $\hat{\theta}_{\hat{K}}$ instead of the full vectors $\hat{\bm{z}}$ and $\hat{\bm{\theta}}$.
Similarly, instead of keeping the full vectors $\hat{\bm{\gamma}}_k$, we keep track only of the priors $\hat{\gamma}_{k,l}<\infty$ and denote their respective indices with index sets $\hat{\mathcal{S}}_k$ for all $k\in\{1,\,2,\,\cdots,\,\hat{K}\}$.
We start with an empty model (bottom-up initialization) where $\hat{K}=0$, $\hat{\bm{\theta}}$, $\hat{\bm{\gamma}}$ and $\hat{\bm{\alpha}}$ are empty vectors, and $\hat{\bm{C}}$ is an empty matrix.
The noise precision is initialized using the signal energy as $\hat{\lambda} = \frac{N}{\|\bm{x}\|^2}$.
Then, we repeatedly alternate between searching a for new group of components to add to the model and updating the already existing groups.
We stop the procedure when the change in $\hat{\bm{x}}=\hat{\bm{\Psi}}\hat{\bm{\alpha}}$ from one iteration to the next is below a threshold and the search does not find a new group to add to the model.

\begin{algorithm}[tbp]
	\renewcommand{\algorithmicrequire}{\textbf{Input:}}
	\renewcommand{\algorithmicensure}{\textbf{Output:}}
	\caption{Main}
	\label{alg:sbcl}
	\begin{algorithmic}
		\REQUIRE Signal vector $\bm{x}$.
		\ENSURE Model order $\hat{K}$, parameters $\hat{\bm{\theta}}$, and amplitudes $\hat{\bm{\alpha}}$.
		\STATE Initialize $\hat{K}=0$, $\hat{\lambda} =\frac{N}{\|\bm{x}\|^2}$, and $\hat{\bm{\alpha}}$, $\hat{\bm{\theta}}$, $\hat{\bm{\gamma}}$ as empty vectors.%
		\REPEAT
		\STATE $\hat{\bm{x}}_{\text{res}}\leftarrow \bm{x}-\hat{\bm{\Psi}}\hat{\bm{\alpha}}$.
		\STATE $\hat{\theta}_{\hat{K}+1} \leftarrow \arg\max_{\theta} |\bm{\psi}^{\hermitian}(\theta,1)\hat{\bm{x}}_{\text{res}}|$.
		\STATE Perform Alg. \ref{alg:fast_prior_updated} to estimate priors $\hat{\bm{\gamma}}_{\hat{K}+1}$ and $\hat{\mathcal{S}}_{\hat{K}+1}$.
		\STATE Calc $\Delta_{\hat{K}+1}(\hat{\theta}_{\hat{K}+1})$ from \eqref{eq:elbo_difference_group}
		\IF{$\Delta_{\hat{K}+1}(\hat{\theta}_{\hat{K}+1})>0$}
		\STATE $\hat{K}\leftarrow \hat{K}+1$.
		\STATE $\hat{\bm{\theta}}\leftarrow \big[\hat{\bm{\theta}}^\transpose,\,\hat{\theta}_{\hat{K}+1}\big]^\transpose,\quad \hat{\bm{\gamma}}\leftarrow \big[\hat{\bm{\gamma}}^\transpose,\,\hat{\bm{\gamma}}_{\hat{K}+1}^\transpose\big]^\transpose$.
		\ENDIF
		\FORALL{groups $k \in \{1,\,2,\,\cdots,\,\hat{K}\}$}
		\STATE Perform Alg. \ref{alg:fast_prior_updated} to update priors $\hat{\bm{\gamma}}_{k}$ and $\hat{\mathcal{S}}_k$.%
		\STATE Find $\hat{\theta}_{k} = \arg\max_{\theta_k} \Delta_k(\theta_k)$ from \eqref{eq:elbo_difference_group}.
		\IF{$\Delta_k(\hat{\theta}_{k}) \le 0$}
		\STATE $\hat{K}\leftarrow \hat{K}-1,\quad \hat{\bm{\theta}}\leftarrow\hat{\bm{\theta}}_{\bar{k}},\quad \hat{\bm{\gamma}} \leftarrow \hat{\bm{\gamma}}_{\bar{k}}$.
		\ENDIF
		\ENDFOR
		\STATE Compute $\hat{\bm{\alpha}}$ and $\hat{\bm{C}}$ from \eqref{eq:update_alpha} and \eqref{eq:update_alphaCovar}.
		\STATE Compute $\hat{\lambda}$ from \eqref{eq:update_lambda}.
		\UNTIL{stopping criterion.}
	\end{algorithmic}
\end{algorithm}

To search for a new group, we would ideally find a combination of $\hat{\theta}_{\hat{K}+1}$ and $\hat{\bm{\gamma}}_{\hat{K}+1}$ which maximizes $\hat{\theta}_{\hat{K}+1} = \arg\max_{\theta} \Delta_{\hat{K}+1}(\theta)$.
Since this is computationally prohibitive, we choose a single component $l\in \mathcal{S}_{\text{max}}$, e.g. $l=1$, and consider a new group parametrized by $\hat{\theta}_{\hat{K}+1}=\arg\max_{\theta} |\bm{\psi}^{\hermitian}(\theta,l)\hat{\bm{x}}_{\text{res}}|$, where $\hat{\bm{x}}_{\text{res}}=\bm{x}-\hat{\bm{\Psi}}\hat{\bm{\alpha}}$ is the residual signal.
Next, we perform Algorithm \ref{alg:fast_prior_updated} to find other related components in the proposed group and calculate the priors $\hat{\bm{\gamma}}_{\hat{K}+1}$ for this new group.
Finally, we add the group to the model if $\Delta_{\hat{K}+1}(\hat{\theta}_{\hat{K}+1})>0$.

\begin{algorithm}[tbp]
	\renewcommand{\algorithmicrequire}{\textbf{Input:}}
	\renewcommand{\algorithmicensure}{\textbf{Output:}}
	\caption{Fast update of priors $\hat{\bm{\gamma}}_k$}
	\label{alg:fast_prior_updated}
	\begin{algorithmic}
		\REQUIRE Signal vector $\bm{x}$, parameters $\hat{\bm{\theta}}, \hat{\bm{\gamma}}, \hat{\mathcal{S}}_k, \hat{\lambda}$, index $k$.
		\ENSURE Prior precisions $\hat{\bm{\gamma}}_k$ and $\hat{\mathcal{S}}_{k}$ of the $k$-th group.
		\FORALL{Compontents $l \in \hat{\mathcal{S}}_k$}
		\STATE $s_{k,l} \leftarrow  (\hat{\lambda}\bm{\psi}_{k,l}^\hermitian \bm{\psi}_{k,l} - \hat{\lambda}^2 \bm{\psi}_{k,l}^\hermitian \bm{\Psi}_{\overline{k,l}}\bm{C}_{\overline{k,l}}\bm{\Psi}_{\overline{k,l}}^\hermitian \bm{\psi}_{k,l})^{-1}$.
		\STATE $u_{k,l} \leftarrow  s_{k,l} (\hat{\lambda} \bm{\psi}_{k,l}^\hermitian\bm{x} - \hat{\lambda}^2 \bm{\psi}_{k,l}^\hermitian \bm{\Psi}_{\overline{k,l}}\bm{C}_{\overline{k,l}}\bm{\Psi}_{\overline{k,l}}^\hermitian\bm{x})$.
		\IF{$\frac{|u_{k,l}|^2}{s_{k,l}} > \chi_1$}
		\STATE $\hat{\gamma}_{k,l} \leftarrow (|u_{k,l}|^2-s_{k,l})^{-1}$.
		\ELSE
		\STATE Remove $\hat{\gamma}_{k,l}$ from $\hat{\bm{\gamma}}_k$.
		\STATE $\hat{\mathcal{S}}_k \leftarrow \hat{\mathcal{S}}_k \backslash \{l\}$.
		\ENDIF
		\ENDFOR
		\FORALL{Compontents $l \in \mathcal{S}_{\text{search}} \backslash \hat{\mathcal{S}}_{k}$}
		\STATE $s_{k,l} \leftarrow (\hat{\lambda}\bm{\psi}_{k,l}^{\hermitian}\bm{\psi}_{k,l} - \hat{\lambda}^2\bm{\psi}_{k,l}^{\hermitian}\hat{\bm{\Psi}}\hat{\bm{C}}\hat{\bm{\Psi}}^{\hermitian}\bm{\psi}_{k,l})^{-1}$.
		\STATE $u_{k,l} \leftarrow s_{k,l}(\hat{\lambda}\bm{\psi}_{k,l}^{\hermitian}\bm{x} - \hat{\lambda}^2\bm{\psi}_{k,l}^{\hermitian}\hat{\bm{\Psi}}\hat{\bm{C}}\hat{\bm{\Psi}}^{\hermitian}\bm{x})$.
		\IF{$\frac{|u_{k,l}|^2}{s_{k,l}} > \chi_1$}
		\STATE $\hat{\bm{\gamma}}_k \leftarrow [\hat{\bm{\gamma}}_k^\transpose,\quad (|u_{k,l}|^2-s_{k,l})^{-1}]^\transpose$.
		\STATE $\hat{\mathcal{S}}_k \leftarrow \hat{\mathcal{S}}_k \cup \{l\}$.%
		\ENDIF
		\ENDFOR
	\end{algorithmic}
\end{algorithm}

After adding a new group, we iterate over all groups $k\in\{1,\,2,\,\cdots,\,\hat{K}\}$ and for each one we first perform an E-step to update the distributions $q_{\gamma,k,l}$, followed by an M-step to update $\hat{\bm{\theta}}_k$ and $\hat{z}_k$.
Lastly, we update the amplitude and noise distributions $q_{\bm{\alpha}}$ and $q_{\lambda}$.
The update of the distributions $q_{\gamma,k,l}$ is outlined in Algorithm \ref{alg:fast_prior_updated} and entails both updating the prior of existing components $\hat{\gamma}_{k,l}$ for all $l\in \hat{\mathcal{S}}_k$ as well as looking for new components to add to the group.
Intuitively, we would calculate $u_{k,l}$ and $s_{k,l}$ for all $l\in \mathcal{S}_{\text{max}}$ to check whether the component should be added or kept in the group or if it should be discarded. However, depending on the application and our choice of $\mathcal{S}_{\text{max}}$ this could be suboptimal. Consider the case of extended object detection. Since we do not want to constrain the size of each object, we are encouraged to use a large range for $\mathcal{S}_{\text{max}}$. However, if two small objects are close to each other this would potentially result in the estimation of only a single group covering both objects with a few spectral lines deactivated in the middle.
To prevent this, we can constrain the search space to $\mathcal{S}_{\text{search}}\subseteq\mathcal{S}_{\text{max}}$ depending on the application. A reasonable choice for the example of extended object detection is to look for new components only in the neighbourhood of the currently existing ones by setting $\mathcal{S}_{\text{search}}=\{\min(\hat{\mathcal{S}}_k)-1,\,\min(\hat{\mathcal{S}}_k),\,\cdots,\,\max(\hat{\mathcal{S}}_k)+1\} \cap  \mathcal{S}_{\text{max}}$.
If such a constrained search space is used, it can be beneficial to run a few updates of each group before adding a new group in order to explore the search space quicker and avoid introducing new groups for components which would be covered by another existing group anyway.

\section{Applications and Results} 
\label{sec:results}

\subsection{Multi-pitch estimation}
Multi-pitch estimation is a fundamental problem in audio signal processing  \cite{benents2019SPM,mueller2011STSP:musicAnalysis,christensen2009:MultiPitchEst,christensen2008SP,adalbjoernsson2015SP,swaerd2018TASLP}.
The goal of multi-pitch estimation is to decompose the signal into several sources, each of which is modeled as a sum of harmonics, giving rise to the harmonically structured model
\vspace*{-1mm}
\begin{align}
	\bm{x} = \sum_{k=1}^{K}\sum_{l\in\mathcal{S}_k} \alpha_{k,l} \ist e^{j2\pi l f_{0,k} \bm{t}} + \bm{\epsilon}\ist.
	\label{eq:multi-pitch-model}\\[-7mm]\nn
\end{align}
Note that \eqref{eq:multi-pitch-model} is an instance of \eqref{eq:signal-model} since the multiples of the fundamental frequencies can be rewritten as $f_{k,l}=\theta_k \ist l$ with $\theta_k=f_{0,k}$.
We aim to estimate the number of sources $K$ along with the fundamental frequency $f_{0,k}$ of each source while $\alpha_{k,l}$ and $\mathcal{S}_{k}$ are considered nuisance parameters.
Even though audio signals are typically real-valued, we can apply the complex-valued signal model by computing the (down-sampled) discrete-time analytical signal \cite{marple1999TSP:analytic-signal}.

To adapt the proposed algorithm to multi-pitch estimation, we refine the search strategy to fit the task at hand.
When looking for new components, we consider all harmonics up to a relative frequency of $\theta_k\ist l=1$.
Thus, we use $\mathcal{S}_{\text{search}}=\{1,\,2,\,\cdots,\,\text{floor}\big(\frac{1}{\hat{\theta}_{k}}\big)\}$.
To find the true fundamental frequency, we also perform a fractional search during which we search for components at fractions $l^\prime \in \{\frac{1}{2},\,\frac{1}{3},\,\cdots,\,\frac{1}{\text{floor}(N \hat{\theta}_{k})}\}$ of the current estimate.
If we find one such component we stop the fractional search and add $l^\prime$ to $\hat{\mathcal{S}}_k$.
In order to obtain integer relations between all components we then re-parametrize $\hat{\theta}_{k}^\prime=\hat{\theta}_{k} \ist l^\prime$ and $\hat{\mathcal{S}}_k^\prime = \frac{\hat{\mathcal{S}}_k}{l^\prime}$.

\subsubsection{Numerical Analysis}
To highlight the robustness of our algorithm against \gls{awgn}, we generate a signal of length $N=100$ samples.
For $N_{\text{MC}}=1000$ simulation runs, the fundamental frequencies of $K=2$ sources with 6 harmonics each are drawn uniformly from the interval $[0.025,\,0.1]$. If the fundamental frequencies are closer than $\frac{2}{N}=0.02$, they are discarded and a new set of fundamental frequencies is drawn. 
We use the \gls{ospa} metric \cite{schuhmacher2008TSP:OSPA} to evaluate the estimation accuracy and cardinality errors of the estimated fundamental frequencies in a single metric. The cutoff-distance for the metric was set to $c=\frac{2}{N}=0.02$ and the order parameter was set to $p=1$.

A uniform prior $p(\theta_k)=\mathcal{U}(\theta_k\,|\,\frac{1}{N},\,1)$ over the full frequency range was applied for the proposed algorithm and the thresholds for component and group sparsity are set to $\chi_1=7\,\text{dB}$ and $\chi_2=10\,\text{dB}$, respectively.
\footnote{We denote the uniform \gls{pdf} as $\mathcal{U}(x|a,b)=\frac{1}{b-a}$ for $a\le x \le b$ and $0$ otherwise.}
We compare our algorithm to the BSURE-IR
\footnote{\url{https://www.maths.lu.se/fileadmin/maths/personal_staff/Andreas_Jakobsson/BSURE.zip}} algorithm \cite{swaerd2018TASLP} and the approximate nonlinear-least-squares (ANLS) \gls{em}
\footnote{\url{https://www.morganclaypool.com/page/multi-pitch}} algorithm of \cite{christensen2008SP}.
The BSURE-IR algorithm was initialized with a grid of $15$ frequency points in the interval $[0.025,\,0.1]$ and the maximum allowed harmonic order was set to $6$. 
The search interval for the ANLS-EM algorithm was set to $[0.025,\,0.1]$ and an FFT size of $2^{12}$ was used.

\begin{figure}
	\centering
	\subfloat[\label{fig:res:multipitch-OSPA-simple}]{\includegraphics{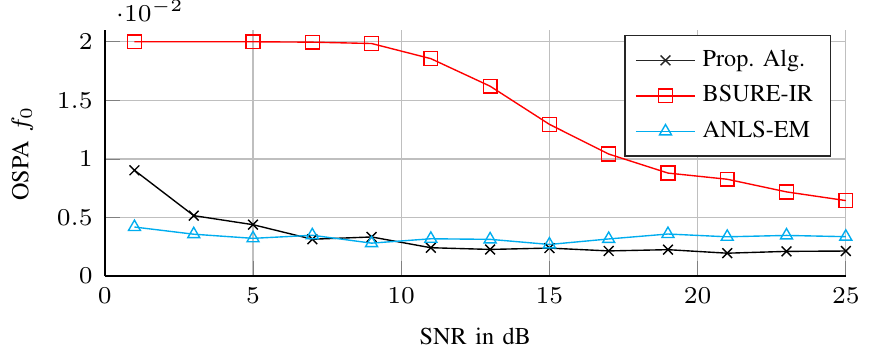}} \\
	\subfloat[\label{fig:res:multipitch-OSPA-tritone}]{\includegraphics{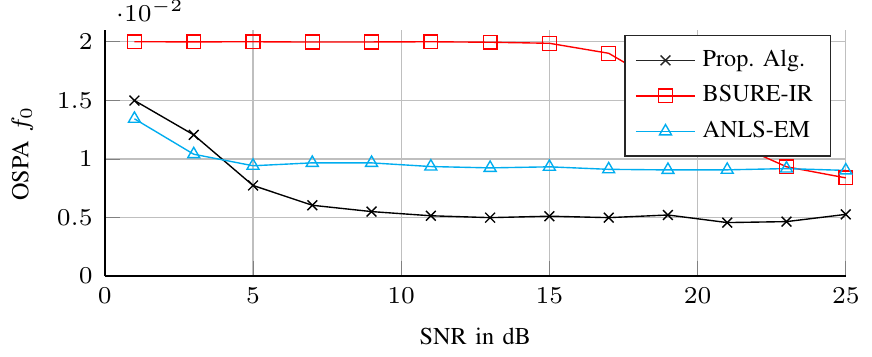}}
	\caption{OSPA of the estimated fundamental frequencies of a signal containing (a) 2 harmonic sources with fundamental frequencies drawn randomly from the interval $[0.025,\,0.1]$ and (b) a major tritone with fundamental frequencies $f_0$, $\frac{5}{4}f_0$ and $\frac{3}{2}f_0$.}
	\label{fig:res:multipitch-OSPA}	
\end{figure}

Figure \ref{fig:res:multipitch-OSPA-simple} shows the  mean \gls{ospa} for all three algorithms versus the \gls{snr} defined as $\text{SNR}=\frac{\|\bm{x}-\bm{\epsilon}\|^2}{\|\bm{\epsilon}\|^2}$.
The BSURE-IR algorithm was not able to find any fundamental frequencies for $\text{SNRs}<10\,\text{dB}$, as indicated by the \gls{ospa} being equal to the cutoff-distance $c$. Even for high \glspl{snr} of $20\,\text{dB}$ and more, the performance of BSURE-IR was worse than that of the proposed algorithm.
The performance of the ANLS-EM algorithm is similar to the proposed algorithm.
For low \glspl{snr} of $5\,\text{dB}$ and less the ANLS-EM outperforms the proposed algorithm in this example.
However, both comparison algorithms have larger average runtime (averaged over simulation runs) than the proposed algorithm.
The average runtime of the BSURE-IR algorithm strongly depends on the \gls{snr}. It is varying from $2.0$ seconds for $\text{SNR} = 20\,\text{dB}$ to $19.3$ seconds for $\text{SNR} = 5\,\text{dB}$.
In contrast, the average runtimes of the proposed algorithm and the ANLS-EM algorithm are approximately constant over the \gls{snr} with $0.1$ seconds for the proposed method and $5.0$ seconds for the ANLS-EM algorithm.

As a second experiment, we generate $N_{\text{MC}}=1000$ simulations runs of a signal consisting of three notes with pitch $f_0$, $\frac{5}{4}f_0$ and $\frac{3}{2}f_0$, i.e. a major triad with perfect temperament and $N=100$ samples.
Each note consists of six harmonics drawn randomly from $\{1,\,2,\,\cdots,\,7\}$.
The fundamental frequency of the base note $f_0$ was drawn uniformly random from the interval $[0.07,\,0.08]$ and all harmonics are generated with unit amplitude and uniformly random phase.
The BSURE-IR algorithm was initialized with 15 grid points in the interval $[0.06,\,0.13]$ while the ANLS-EM algorithm was given the same range as frequency prior and the FFT-size was again set to $2^{12}$.
The settings for the proposed algorithm are the same as in the previous experiment.
Figure \ref{fig:res:multipitch-OSPA-tritone} shows again the mean \gls{ospa} versus the \gls{snr}.
In this scenario the proposed algorithm performs better than ANLS-EM algorithm for almost all \gls{snr} values.
This is mainly due to the fact that the ANLS-EM algorithm underestimates the model order in most cases, even for high \gls{snr} values of $20\,\text{dB}$ and more.
This indicates a performance degradation of the ANLS-EM algorithm in more complex scenarios containing more fundamental frequencies and overlapping harmonics. Again, The performance of BSURE-IR is the worst of the three, even in the high \gls{snr} regime.

\subsubsection{Real Music Signals}
\begin{figure}
	\centering
	\includegraphics{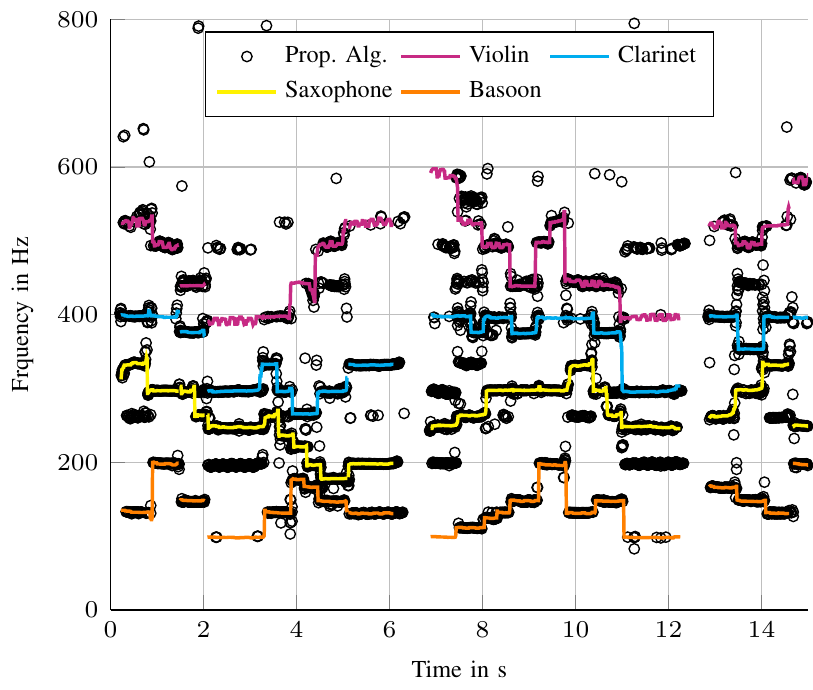}
	\caption{Pitch estimates for $15$ seconds of the choral \emph{Ach Gott und Herr} compared to the ground truth (best viewed in color). The performance of the proposed algorithm for the depicted timespan is $\text{accuracy}=0.55$, $\text{precision}=0.75$ and $\text{recall}=0.68$.}
	\label{fig:res:bach10Example}
	\vspace*{-1mm}
\end{figure}

To evaluate the performance of the proposed algorithm on real data, we apply it to estimate the fundamental pitches in the Bach-10 dataset
\footnote{\url{https://labsites.rochester.edu/air/resource.html}} \cite{duan2010TASLP:Bach10}.
The dataset contains 10 chorales of J.S. Bach played by a quartet consisting of a violin, a clarinet, a saxophone and a bassoon.
Each piece lasts between 25-40 seconds with all instruments playing nearly all the time.
The audio of each instrument was recorded individually while the musician listened to the others via headphones. The fundamental pitch of each instrument was extracted from the individual recordings using the YIN single pitch estimator \cite{cheveigne2002JASA:YIN} as ground truth.
Obvious errors in the ground truth are corrected manually.
The audio signal was segmented into frames of $45\,\text{ms}$ with a $10\,\text{ms}$ stride between frames and the proposed algorithm was applied to each frame individually.
Since the audio quality is quite good, we select the thresholds to be $\chi_1=15\,\text{dB}$ and $\chi_2=21,\,\text{dB}$, respectively, and applied a uniform prior between $75\,\text{Hz}$ and $10\,\text{kHz}$.
Each pitch in the ground truth was considered matched if an estimated pitch deviated from it no more than a halve of a semitone, i.e. if it deviates no more than $3\,\%$ from the true pitch.
Let $\text{TP}^{(i)}$ be the number of pitches matched between the ground truth and the estimate in each frame $i\in\{1,\,2,\,\cdots,\,N_{\text{Frames}}\}$, $\text{FP}^{(i)}$ the number of false positives, i.e. the number of estimated pitches which did not match to a ground truth pitch, and $\text{FN}^{(i)}$ the number of ground truth pitches which were not matched to an estimate, and $N_{\text{Frames}}$ the total number of frames.
Table \ref{tab:res:multiPitch-Bach10} lists the accuracy, precision, recall and the $\text{F}_1$ of the proposed algorithm along with several comparison methods. The values were calculate using \cite{bay2009ISMIR}
	\begin{align}
		\text{Accuracy} &= \frac{\sum_{i=1}^{N_{\text{Frames}}} \text{TP}^{(i)}}{\sum_{i=1}^{N_{\text{Frames}}} \text{TP}^{(i)} + \text{FP}^{(i)} + \text{FN}^{(i)}} \\		
		\text{Precision} &= \frac{\sum_{i=1}^{N_{\text{Frames}}} \text{TP}^{(i)}}{\sum_{i=1}^{N_{\text{Frames}}} \text{TP}^{(i)} + \text{FP}^{(i)}} \\
		\text{Recall} &= \frac{\sum_{i=1}^{N_{\text{Frames}}} \text{TP}^{(i)}}{\sum_{i=1}^{N_{\text{Frames}}} \text{TP}^{(i)} + \text{FN}^{(i)}} \\
		\text{F}_{1} &= \frac{2 \cdot \text{Precision}\cdot \text{Recall}}{\text{Precision} + \text{Recall}}
		.
	\end{align}
The comparison methods are two model based methods, BSURE-IR \cite{swaerd2018TASLP} and PEARLS \cite{elvander2017TASLP}, as well as a pretrained method \cite{benetos2015ISMIR} denote here as BW15. The results for BW15 are take from \cite{elvander2017TASLP} since they reported an overall better accuracy for the method compared to the original paper.
The ANLS-EM was not included in the table, since preliminary investigations showed a significantly worse performance on the dataset than the other algorithms, highlighting again the inability of the algorithm to cope with more complicated scenarios.
Figure \ref{fig:res:bach10Example} shows the ground truth as well as the estimated fundamental frequencies obtained by the proposed algorithm for several frames of the choral \emph{``Ach Gott und Herr''} of the dataset.
The proposed algorithm is able to capture most of the fundamental frequencies with very few false positives and it outperforms all three state-of-the-art comparison methods, even BW15 that is pre-trained on the instruments in the dataset.

\begin{table}
	\centering
	\caption{Performance measures for the proposed algorithm evaluated on the Bach-10 dataset}
	\label{tab:res:multiPitch-Bach10}
	\begin{tabular}{lccccc}
		Method & $\text{F}_{1}$ & Accuracy & Precision & Recall & Pre-Trained \\ \hline
		Prop. Alg. & \textbf{0.72} & \textbf{0.56} & \textbf{0.73} & \textbf{0.70} & No \\
		BW15 & 0.67 & 0.52 & 0.68 & 0.68 & Yes \\
		BSURE-IR & 0.64 & 0.47 & 0.68 & 0.54 & No \\
		PEARLS & 0.60 & 0.44 & 0.56 & 0.51 & No 
	\end{tabular}
\end{table}

We considered only methods which work on the basis of individual frames to keep the comparison fair.
Naturally, there exist several methods for multi-pitch estimation which not only estimate the pitches on a frame-by-frame basis but track notes over multiple frames to increase the performance. 
Thereby, these methods are able to achieve an $\text{F}_1$ score of up to $0.85$ on the Bach 10 dataset. See e.g. \cite{li2022CC} and references therein.
However, these methods are usually data-driven and specifically tailored to multi-pitch estimation whereas our algorithm is the best purely model-based algorithm without being specifically tailored to the problem at hand. Furthermore, we also expect the proposed algorithm to generalize better to new datasets, out-of-domain data or noisy input data than data-driven approaches.
Additionally, we expect that the performance of the proposed algorithm can be increased significantly by fusing information between frames.

\subsection{Extended Object Detection}

A well-studied problem in radar signal processing is the detection of extended objects \cite{granstrom2017JAIF:EO,schubert2013TAP}. An extended object is defined as a volume over which scatter points are distributed which correspond to a single physical target such as a car. The transmitted signal $s(t)$ is reflected by each of $L_k$ scatter points of the $k$-th target and reaches the receiving antenna with some delay $\tau_{k,l}>0$ with some amplitude $\alpha_{k,l} \in \mathbb{C}$. Thus, the received baseband signal can be modeled as
\vspace*{-1mm}
\begin{align}
	r(t) = \sum_{k=1}^{K} \sum_{l=1}^{L_k} \alpha_{k,l} s(t-\tau_k) + \epsilon(t)\ist.
	\label{eq:extended-object-original-time}\\[-6mm]\nn
\end{align}
Let $\Delta f=\frac{1}{NT_{\text{s}}}$, $R(f)$ be the Fourier transform of $r(t)$ and $\bm{r} = [R(-\frac{N}{2}\Delta f)\iist R((-\frac{N}{2}+1)\Delta f)\,\cdots\,R((\frac{N}{2}-1)\Delta f)]^\transpose$.
Similarly, let $S(f)$ be the Fourier transform of $s(t)$, $\bm{s}=[S(-\frac{N}{2}\Delta f)\iist S((-\frac{N}{2}+1)\Delta f)\,\cdots\,S((\frac{N}{2}-1)\Delta f)]^\transpose$ and $\bm{\epsilon}$ be a sampled vector of the noise in frequency domain. Furthermore, let $\bm{f}=[-\frac{N}{2}\Delta f \iist (-\frac{N}{2}+1)\Delta f \,\cdots \,(\frac{N}{2}-1)\Delta f]^\transpose$.
We can then express \eqref{eq:extended-object-original-time} in the frequency domain as
\vspace*{-1mm}
\begin{align}
	\bm{r} = \sum_{k=1}^{K} \sum_{l=1}^{L_k} \alpha_{k,l} \bm{s}\odot e^{-j2\pi \bm{f} \tau_{k,l}} + \bm{\epsilon}\ist.
	\label{eq:extended-object-original-freq}\\[-6mm]\nn
\end{align}
If the pulse is sufficiently short in time domain, we can apply the sampling theorem to approximate \eqref{eq:extended-object-original-time} with a few signal samples (``tabs'') spaced with delay $\Delta \tau=\frac{1}{f_{\text{s}}}$. Thus, \eqref{eq:extended-object-original-freq} is well approximated by
\vspace*{-1mm}
\begin{align}
	\bm{r} = \sum_{k=1}^{K} \sum_{l=0}^{L_k^\prime} \big(\bm{s}\odot e^{-j2\pi \bm{f} (\tau_{0,k}+l\Delta \tau)}\big)\alpha_{k,l} + \bm{\epsilon}
	\label{eq:extended-object-model}\\[-6mm]\nn
\end{align}
where $\tau_{0,k}=\min_l(\tau_{k,l})$ is the smallest delay of the $k$-th target signal.
Since the model \eqref{eq:extended-object-model} is an instance of \eqref{eq:signal-model}, the proposed algorithm can be applied for the detection and estimation of the radar response from extended objects using $\theta_k=\tau_{0,k}$, $\bm{\psi}(\tau_{0,k},l)=\bm{s}\odot e^{-2j\pi\bm{f}\tau_{k,l}}$ and $\tau_{k,l} = \tau_{0,k}+l\Delta\tau$.

In order to showcase the ability of the algorithm to detect weak object signals and estimate their properties, we consider the case of a single object in noise and run a numerical experiment of $N_{\text{MC}}=10^5$ realizations.
The object is modeled according to \cite{schubert2013TAP} with a uniform intensity function $q(\tau)$ such that on average 10 scatter point are drawn between $\tau=32.323\, T_{\text{s}}$ to $\tau=37.323\, T_{\text{s}}$. Each scatter point was modeled to have an amplitude $\alpha_{k,l} \sim \mathcal{CN}(\alpha_{k,l}|0,\,1)$ drawn independently from a complex Gaussian distribution with zero mean and unit variance.

\begin{figure}
	\centering
	\subfloat[\label{fig:res:eo-rmse-cdf}]{\includegraphics{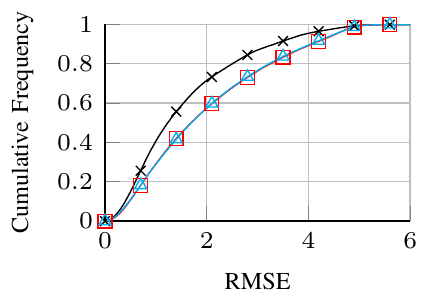}}
	\hfill
	\subfloat[\label{fig:res:eo-rmse-snr}]{\includegraphics{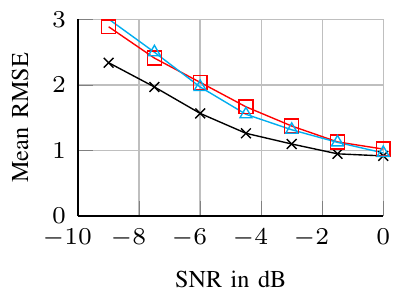}}
	\\[-1mm]
	\subfloat[\label{fig:ref:eo-histogram}]{\includegraphics{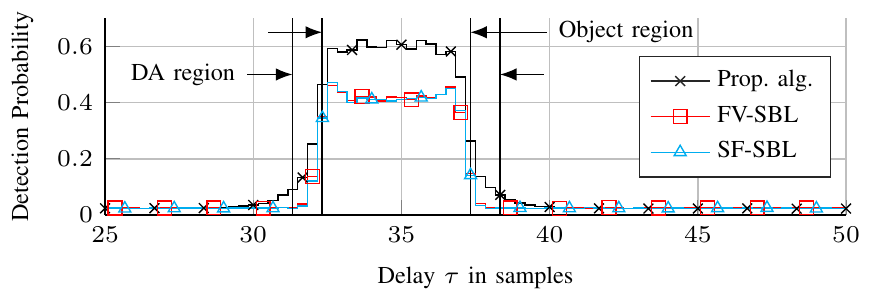}}
	\caption{Comparison of the proposed algorithm compared to the FV-\gls{sbl} \cite{shutin2013:VSBL} and SF-\gls{sbl} \cite{hansen2018TSP:SuperFastLSE}. Cumulative frequency of the RMSE at an SNR of $-6\,\text{dB}$ (a), mean RMSE over SNR (b) and histogram of the delay of the detected components at an SNR of $-6\,\text{dB}$ (c).}
	\label{fig:res:extended-object}
	\vspace*{-1mm}
\end{figure}

To apply the proposed algorithm to the problem, we select $\mathcal{S}_{\text{search}}=\{\min(\hat{\mathcal{S}_k})-1,\,\min (\hat{\mathcal{S}_k}),\,\cdots,\,\max(\hat{\mathcal{S}_k})+1\}$ and set the threshold to $\chi_1=3\,\text{dB}$ and $\chi_2=4\,\text{dB}$.
As comparison algorithms, we use two variants of \gls{sbl}-based superresolution algorithms, the algorithm proposed in \cite{shutin2013:VSBL} abbreviated FV-SBL and the algorithm proposed in \cite{hansen2018TSP:SuperFastLSE} abbreviated SF-SBL. Note that the SF-SBL is similar to the proposed algorithm using $\mathcal{S}_{\text{max}}=\{0\}$ (i.e. allowing only a single component per group) and $\chi_1=0\,\text{dB}$.
To make the comparison fair, we selected the thresholds of the three algorithms such the mean number of components estimated outside the object region, was approximately the same.
Numerical analysis revealed that this is achieved by thresholds of $6\,\text{dB}$ for the FV-SBL and $\chi_2=3.5\,\text{dB}$ for the SF-SBL.
To estimate the extent $\hat{O}_{\text{E}}$ and center-of-mass $\hat{O}_{\text{C}}$ of the object we used an ``oracle'' data association which considered all components within $\pm1$ sample of the true object region to belong to the object and ignored the grouping estimated by the proposed algorithm.
This data association is intended to reflect the information which can be obtained in case the algorithm is used to preprocess measurements for an extended object tracking filter which performs the data association.
Let $\hat{\alpha}_{\text{o},l}$ and $\hat{\tau}_{\text{o},l}$ for $l\in\{1,\,2,\,\cdots,\,\hat{L}\}$ denote respective amplitudes and delays associated with the object, we estimated $\hat{O}_{\text{E}}$ and $\hat{O}_{\text{C}}$ as
\begin{align}
	\hat{O}_{\text{E}} &= \max_l(\hat{\tau}_{\text{o},l})-\min_l(\hat{\tau}_{\text{o},l}) \label{eq:extend}\\
	\hat{O}_{\text{C}} &=\frac{\sum_{l=1}^{\hat{L}}|\hat{\alpha}_{\text{o},l}|^2\hat{\tau}_{\text{o},l}}{\sum_{l=1}^{\hat{L}}|\hat{\alpha}_{\text{o},l}|^2}\label{eq:center}\ist.
\end{align}
Using \eqref{eq:extend} and \eqref{eq:center}, the root mean squared error (RMSE) is given by $\text{RMSE} = \sqrt{(\hat{O}_{\text{E}}-O_{\text{E}})^2 + (\hat{O}_{\text{C}}-O_{\text{C}})^2}$ where $O_{\text{E}}$ and $O_{\text{C}}$ are the true object extent and center-of-mass.
The cumulative frequency of the RMSE for an \gls{snr} of $\frac{\|\bm{r}-\bm{\epsilon}\|^2}{\|\bm{\epsilon}\|^2}=-6\,\text{dB}$ is depicted in Figure \ref{fig:res:extended-object}a while the mean RMSE over SNR is depicted in Figure \ref{fig:res:extended-object}b.
The proposed algorithm is able to outperform both SBL-based superresolution methods in terms of RMSE at all \gls{snr} levels, although the difference is more pronounced in low-\gls{snr} conditions of $-3\,\text{dB}$ and less.
The reason for this becomes evident when investigating the histogram of the locations of detected components depicted in Figure \ref{fig:res:extended-object}c.
While the mean number of components estimated in the object region is similar for the comparison methods, the proposed algorithms detects more components in the object region and is therefore able to estimate the parameters of the object more accurately.
The increased number of estimated components is due to the lower threshold when adding new components within a group compared to adding new groups.
Additional investigations not included in the paper revealed that the RMSE difference between the proposed algorithm and the comparison methods stems mainly from a more accurate estimation of the extent of the object while the estimation performance for the center-of-mass is similar for all three algorithms.

\subsection{Variational Mode Decomposition}
Another task which can be (approximately) solved by the proposed algorithm is \gls{vmd} \cite{dragomiretskiy2014TSP:VMD}. \Gls{vmd} decomposes a signal $\bm{x}$ into several ``intrinsic mode functions'' $\bm{x_k}$
\vspace*{-1mm}
\begin{align}
	\bm{x} = \sum_{k=1}^{K} \bm{x_k}\ist.
	\label{eq:vmd-original}\\[-6mm]\nn
\end{align}
An ``intrinsic mode function'' $x_k(t)=A_k(t)\cos(\varphi_k(t))$ is defined in \cite{dragomiretskiy2014TSP:VMD} as a sinusoidal function where the amplitude $A_k(t) \ge 0$ changes slowly over time $t$ and the phase $\varphi_k(t)$ is a non-decreasing function with slowly time-varying instantaneous frequency $\frac{d\varphi_k(t)}{dt}$.
Furthermore, we consider the discrete signal  $\bm{x_k}=[x_k(-\frac{N}{2}T_{\text{s}})\iist x_k((-\frac{N}{2}+1)T_{\text{s}})\,\cdots\,x_k((\frac{N}{2}-1)T_\text{s})]^\transpose$ sampled with regular intervals  $f_{\text{s}}=\frac{1}{T_{\text{s}}}$.
The model \eqref{eq:vmd-original} is not an instance of \eqref{eq:signal-model}.
First, it considers real signals $\bm{x}$ and $\bm{x}_k$ instead of complex ones.
However, This can be sidestepped by computing the discrete time analytical signal \cite{marple1999TSP:analytic-signal}.
Secondly, \eqref{eq:vmd-model} does not consider the signal to be embedded in additive noise. However, we find that modelling the noise is beneficial since noise is present in many practical applications anyway.
Another difference between \gls{vmd} and the algorithm developed here is, that \gls{vmd} assumes the number of modes $K$ is known a priori whereas the developed algorithm estimates $K$.
Finally, it may not be immediately clear how to model each $\bm{x}_k$ as a discrete sum of components.
Since $x_k(t)$ is essentially an amplitude and phase modulated signal, almost all of the energy of the signal will be within a bandwidth which is much smaller than the sampling rate $f_{\text{s}}$.
Therefore, each $x_k(t)$ can be approximated by a signal with finite bandwidth which, by the sampling theorem, can be represented as a set of frequency samples (``tabs'') spaced with $\Delta f=\frac{f_{\text{s}}}{N}$.
Thus, we approximate the discrete-time analytic signal of each mode as $\bm{x}_{\text{a},k} \approx \sum_{l=-L_k}^{L_k}\alpha_{k,l} \ist e^{j2\pi f_{k,l}\bm{t}}$, where $f_{k,l} = \theta_k+l\Delta f$ and $L_k$ relates to the bandwidth $B_k$ of the $k$-th mode.
The analytic signal $\bm{x}_{\text{a}}$ is modeled as
\vspace*{-1mm}
\begin{align}
	\bm{x}_{\text{a}} \approx \sum_{k=1}^{K} \sum_{l=-L_k}^{L_k} \alpha_{k,l} \ist e^{j2\pi f_{k,l}\bm{t}} + \bm{\epsilon}
	\label{eq:vmd-model}\\[-6mm]\nn
\end{align}
which is an instance of \eqref{eq:signal-model}.
The original signals (or estimates thereof) can be obtained as the real part of the corresponding analytical signal.
We used a larger search radius of $\mathcal{S}_{\text{search}}=\{\min(\hat{\mathcal{S}}_k)-5,\,\min(\hat{\mathcal{S}}_k)-4,\,\cdots,\,\max(\hat{\mathcal{S}}_k)+5\}$ due to the larger signal length and set $\chi_1=3\,\text{dB}$ as parameters for the proposed method. We set $\chi_2=10\,\text{dB}$ and $\chi_2=18\,\text{dB}$ for the $10\,\text{dB}$ \gls{snr} and $30\,\text{dB}$ \gls{snr} case, respectively, based on preliminary investigations.

\begin{figure}
	\centering
	\subfloat[\label{fig:res:vmd-examples}]{\includegraphics{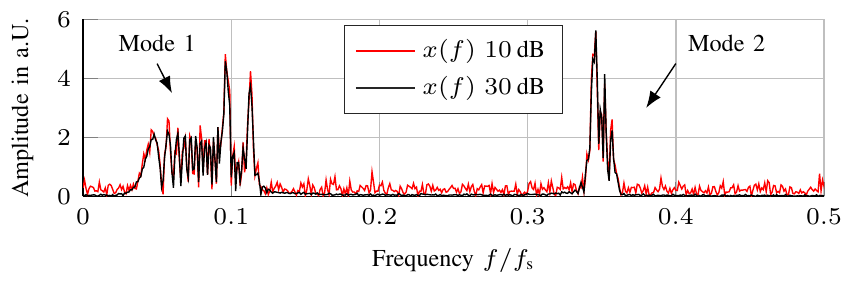}}
	\\
	\includegraphics{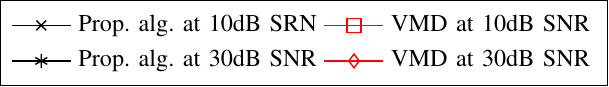} \\[-3mm]
	\subfloat[\label{fig:res:vmd-cdf-1}]{\includegraphics{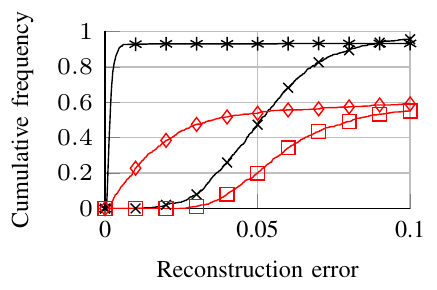}}	
	\hfill
	\subfloat[\label{fig:res:vmd-cdf-2}]{\includegraphics{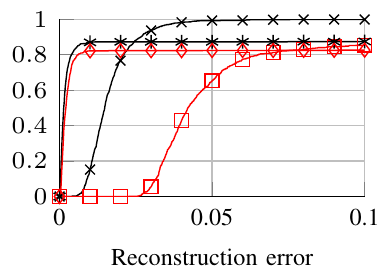}}
	\caption{Performance of the proposed algorithm for the task of \gls{vmd}. (a) Example signal consisting of two modes, (b) cumulative frequency for the reconstruction error of mode 1 and (c) mode 2.}
	\label{fig:res:vmd-accuracy}
	\vspace*{-1mm}
\end{figure}

To demonstrate the estimation accuracy of the underlying modes, we generate a signal of length $N=1000$ samples consisting of two modes in \gls{awgn} according to \eqref{eq:vmd-original}.
The amplitude and instantaneous frequency are defined at support points $t_m\in \{-\frac{N}{2}T_{\text{s}},\,-\frac{N}{4}T_{\text{s}},\,0,\,\frac{N}{4}T_{\text{s}},\,(\frac{N}{2}-1)T_{\text{s}}\}$ and linearly interpolated in between. The amplitude support points are defined as $A_{k,m}=A_{k,0}(1+A_{\text{mod},k,m})$ where $A_{\text{mod},k,m}\sim\mathcal{U}(-A_{\text{mod},k}, A_{\text{mod},k})$ is a uniform random variable drawn independently for each mode $k$ and support point $m$. Similarly, the instantaneous frequency at $t_m$ is defined as  $\frac{d\varphi_k(t)}{dt}|_{t=t_m} = 2\pi f_k (1+f_{\text{mod},k,m})$ and linearly interpolated in between the points, where $f_{\text{mod},k,m}\sim \mathcal{U}(-f_{\text{mod},k},f_{\text{mod},k})$ is again an i.i.d. uniform random variable. The phase $\varphi_k[n]$ is obtained by integrating the instantaneous frequency from $t=-\frac{N}{2} T_{\text{s}}$ to $nT_{\text{s}}$.
We select the modulation parameters for the first mode to be $f_k=0.1$, $f_{\text{mod},k}=0.66$, $A_{k,0}=1$ and $A_{\text{mod},k}=0.25$ and for the second mode we select $f_k=0.35$, $f_{\text{mod},k}=0.03$, $A_{k,0}=1$ and $A_{\text{mod},k}=0.9$.
Thus, the amplitude modulation is more pronounced in the second mode and the frequency modulation is more pronounced in the first mode. An example signal resulting from these settings is shown in Figure \ref{fig:res:vmd-accuracy}a.

We run a numerical simulation with $N_{\text{MC}}=1000$ realizations of \eqref{eq:vmd-original} with added real valued Gaussian noise $\bm{\epsilon}$ of an \gls{snr} of $\frac{\|\bm{x}\|^2}{\|\bm{\epsilon}\|^2}=10\,\text{dB}$ representing a noisy signal and an \gls{snr} of $30\,\text{dB}$ representing a (nearly) noiseless signal.
We apply the proposed algorithm to the analytical signal and compare the estimation accuracy of the modes to the Matlab implementation (release R2022b) of the \gls{vmd} algorithm.
As performance metric we compared the estimation accuracy 
\vspace*{-1mm}
\begin{align}
	\hat{e}_k = \min_l \frac{\|\bm{x}_k - \hat{\bm{x}}_l\|^2}{\|\bm{x}_k\|^2}\\[-6mm]\nn
\end{align}
where $\bm{x}_k$ are the true modes of the signal $\bm{x}$ and $\hat{\bm{x}}_l$ are the estimated modes.
Figures \ref{fig:res:vmd-accuracy}b and \ref{fig:res:vmd-accuracy}c depict the cumulative frequencies of the estimation error of both modes.
The estimation performance of the proposed algorithm is better than the estimation performance of the \gls{vmd} algorithm for both modes.
However, it should be noted that the cumulative frequency of the estimation error ``flattens out'' at different values instead of converging towards 1, indicating some problem with the estimation process.
Since the \gls{vmd} algorithm was designed for narrowband modes, one reason for the poor estimation performance of the \gls{vmd} algorithm on Mode 1 could be the large bandwidth of this mode. However, the proposed method was able to outperform the \gls{vmd} algorithm even for the narrower second mode.
The proposed algorithm overestimated the model order in $5\,\%$ and $19\,\%$ of the time for the $10\,\text{dB}$ \gls{snr} case and $30\,\text{dB}$ \gls{snr} case, respectively.
In these cases, an additional mode appears, resulting in a significantly decreased estimation performance, which leads to the cumulative error frequency plateauing at a value $<1$.

Although we leave an extensive investigation of the performance of the proposed algorithm for \gls{vmd} for future research, we chose to include these results to showcase how the algorithm can be applied to different problem settings.

\section{Conclusion}
\label{sec:conclusion} 

We derive an algorithm for the estimation of structured line spectra.
Such structured line spectra can be found in many different fields, such as multi-pitch estimation, extended object detection using radar signals or \gls{vmd}.
Our algorithm is based on variational Bayesian inference and a Bernoulli-Gamma-Gaussian hierarchical prior model.
In this model, the occurrence of groups is regularized by a Bernoulli prior while the occurrence of spectral lines is regularized by a Jeffrey's prior on the amplitude variances.
Thus, a sparse estimate is obtained which automatically estimates the model order and the group structure in addition to the group parameters and component amplitudes.
Due to the Bernoulli prior, the model is also more resilient to the insertion of additional artificial components compared to Gamma-Gaussian model used typically in \gls{sbl}.
The model can be straightforwardly adapted to a variety of inference problems based on the relation between the component frequencies $f_{k,l}$ and the group parameter $\theta_k$.

We demonstrate the superior performance of our algorithm compared to state-of-the art algorithms for multi-pitch estimation using numerical simulations and show that it is able to outperform state-of-the art multi-pitch estimation algorithms when estimating the fundamental pitches in a major triad.
Additionally, the developed algorithm is shown to outperform state-of-the-art model-based as well as pre-trained algorithms for multi-pitch estimation in accuracy, precision and recall when evaluated on the Bach-10 dataset.
As a second example, we demonstrated how exploiting the structure of the underlying line spectra can be used to increase the estimation performance for the extent and center-of-mass of an extended object in low-\gls{snr} conditions compared to \gls{lse} algorithms for unstructured line spectra. As a third example, we detail how the proposed algorithm can be adapted to perform \gls{vmd} and our algorithm is shown to outperform the Matlab implementation of the \gls{vmd} algorithm in the presented example. However, a more extensive characterization of the performance of \gls{vmd} should be performed by future research.
These three examples demonstrate the versatility of the developed algorithm and show how integrating knowledge about the structural relations between spectral lines into the estimation procedure can lead to performance gains in the low \gls{snr} regime.

Promising directions for future research include the extension of the proposed algorithm to sequential processing based on belief propagation message passing \cite{xuhong2022TWC}, such as for radio-frequency simultaneous localization and mapping \cite{LeitMeyHlaWitTufWin:TWC2019,LeiVenTeaMey:Arxiv2022}, as well as using variational autoencoders \cite{johnson2016NeurIPS:SVAE} to learn arbitrary structured dictionaries.

\appendix
\subsection{Derivation of the Variational Update Equations}
\label{sec:appendix:update-equations}

For the derivation of $q_{\bm{\alpha}}$ we insert \eqref{eq:posterior-distribution} into \eqref{eq:message_update}.
After taking the logarithm on both sides and ignoring all terms which do not depend on $\bm{\alpha}$, we get
\begin{align}
	\ln q_{\bm{\alpha}}&(\bm{\alpha};\hat{\bm{\theta}},\hat{\bm{z}})  \nonumber\\ 
	&\overset{e}{\propto}\big<\ln p(x|\bm{\alpha},\bm{\theta}=\hat{\bm{\theta}}) + \ln p(\bm{\alpha}|\bm{\gamma},\bm{z}=\hat{\bm{z}})\big>_{q_{\lambda}q_{\bm{\gamma}}}\label{eq:AppEq1}.
\end{align}
Inserting \eqref{eq:prior-alpha} and \eqref{eq:likelihood} into \eqref{eq:AppEq1} and ignoring all terms which do not depend on $\bm{\alpha}$, \eqref{eq:AppEq1} can be rewritten as
\begin{align}
	\ln q_{\bm{\alpha}} &\overset{e}{\propto} \big<-\lambda(\bm{x}-\hat{\bm{\Psi}}\bm{\alpha}_{\hat{\mathcal{S}}})^\hermitian(\bm{x}-\hat{\bm{\Psi}}\bm{\alpha}_{\hat{\mathcal{S}}}) - \bm{\alpha}_{\hat{\mathcal{S}}}\bm{\Gamma}\bm{\alpha}_{\hat{\mathcal{S}}} \nonumber \\
	&\qquad+ \sum_{\alpha_{k,l}\notin \bm{\alpha}_{\hat{\mathcal{S}}}} \ln \delta(\alpha_{k,l})\big>_{q_{\lambda}q_{\bm{\gamma}}}
	\label{eq:AppEq2}.
\end{align}
Using the expectation of $q_{\lambda}$, i.e., $\hat{\lambda}=\big<\lambda\big>_{q_{\lambda}}$ (see \eqref{eq:update_lambda}) and the mean of $q_{\gamma,k,l}$, i.e., $\hat{\gamma}_{k,l}=\big<\gamma_{k,l}\big>_{q_{\gamma,k,l}}$ (see \eqref{eq:gamma_convergence_single}) and denoting the real operator as $\Re\{\cdot\}$, \eqref{eq:AppEq2} can be rewritten
\begin{align}
\hspace*{-1.5mm}\ln q_{\bm{\alpha}}&\overset{e}{\propto} -\big(\bm{\alpha}_{\hat{\mathcal{S}}}^\hermitian(\hat{\lambda}\hat{\bm{\Psi}}^\hermitian\hat{\bm{\Psi}} + \hat{\bm{\Gamma}})\bm{\alpha}_{\hat{\mathcal{S}}} - \Re\{\bm{\alpha}_{\hat{\mathcal{S}}}^\hermitian 2\hat{\lambda}\hat{\bm{\Psi}}^\hermitian\bm{x}\} \big) \nonumber \\
&\qquad + \sum_{\alpha_{k,l}\notin\bm{\alpha}_{\hat{\mathcal{S}}}} \ln \delta(\alpha_{k,l}) \nonumber\\	&\overset{e}{\propto}-(\bm{\alpha}_{\hat{\mathcal{S}}}-\hat{\bm{\alpha}})^\hermitian\hat{\bm{C}}^{-1}(\bm{\alpha}_{\hat{\mathcal{S}}}-\hat{\bm{\alpha}}) + \hspace*{-1.5mm}\sum_{\alpha_{k,l}\notin\bm{\alpha}_{\hat{\mathcal{S}}}}\hspace*{-1mm}\ln\delta(\alpha_{k,l}).\\[-7mm]\nonumber
\end{align}
After ``completing the squares'' using $\hat{\bm{C}}=(\hat{\lambda}\hat{\bm{\Psi}}^\hermitian\hat{\bm{\Psi}} + \hat{\bm{\Gamma}})^{-1}$ and $\hat{\bm{\alpha}}=\hat{\lambda}\hat{\bm{C}}\hat{\bm{\Psi}}^\hermitian\bm{x}$ we arrive at \eqref{eq:q-alpha}, \eqref{eq:update_alpha} and \eqref{eq:update_alphaCovar} with $q_{\bm{\alpha}}$ being a complex Gaussian distribution of $\bm{\alpha}_{\hat{\mathcal{S}}}$ and $\alpha_{k,l}=0$ for $\alpha_{k,l}\notin \bm{\alpha}_{\hat{\mathcal{S}}}$.

For the update of $q_{\lambda}$, we start again by inserting \eqref{eq:posterior-distribution} into \eqref{eq:message_update}, applying the logarithm on both sides, and ignoring all terms which do not depend on $\lambda$, i.e., 
\begin{equation}
	\ln q_{\lambda}(\lambda;\hat{\bm{\theta}},\hat{\bm{z}}) \overset{e}{\propto} \big<\ln p(\bm{x}|\bm{\alpha},\bm{\theta}=\hat{\bm{\theta}}) + \ln p(\lambda)\big>_{q_{\bm{\alpha}}q_{\bm{\gamma}}}
	\label{eq:AppEq3}.
\end{equation}
Inserting the likelihood \eqref{eq:likelihood} and $p(\lambda)=\text{Ga}(\lambda|\rho,\mu)$ into \eqref{eq:AppEq3}, we get
\begin{align}
\hspace*{-1mm}\ln p_{\lambda} &\overset{e}{\propto}  \big<\ln |\pi \lambda^{-1}\bm{I}| - \lambda(\bm{x}-\hat{\bm{\Psi}}\bm{\alpha}_{\hat{\mathcal{S}}})^\hermitian(\bm{x}-\hat{\bm{\Psi}}\bm{\alpha}_{\hat{\mathcal{S}}}) \nonumber \\
	&\qquad + (\rho-1)\ln \lambda - \lambda \mu \big>_{q_{\bm{\alpha}}q_{\bm{\gamma}}}\nonumber \\
	&\overset{e}{\propto} (N+\rho-1)\ln \lambda - \lambda\big(\big<\|\bm{x}-\hat{\bm{\Psi}}\bm{\alpha}_{\hat{\mathcal{S}}}\|^2\big>_{q_{\bm{\alpha}}} + \mu \big)
	.
\end{align}
After solving the expectation of $\|\bm{x}-\hat{\bm{\Psi}}\bm{\alpha}_{\hat{\mathcal{S}}}\|^2$ over $q_{\bm{\alpha}}$, we find that $q_{\lambda}$ is again a Gamma distribution with shape $N+\rho$ and rate $\|\bm{x}-\hat{\bm{\Psi}}\hat{\bm{\alpha}}\|^2 + \text{tr}(\hat{\bm{\Psi}}\hat{\bm{C}}\hat{\bm{\Psi}}^\hermitian) + \mu$, as in \eqref{eq:q-lambda} and \eqref{eq:update_lambda}.

For the derivation of $q_{\gamma,k,l}$ given that $\hat{z}_k=1$, we again start with
\begin{align}
	\ln q_{\gamma,k,l}(\gamma_{k,l};\hat{\bm{\theta}},\hat{\bm{z}}) &\overset{e}{\propto}\big<\ln p(\bm{\alpha}|\bm{\gamma},\bm{z}=\hat{\bm{z}}) + \ln p(\gamma_{k,l})\big>_{q_{\bar{k}}} \nonumber \\
	&\overset{e}{\propto} \ln |\pi \bm{\Gamma}^{-1}|^{-1} -\big<\bm{\alpha}_{\hat{\mathcal{S}}}^\hermitian\bm{\Gamma}\bm{\alpha}_{\hat{\mathcal{S}}}\big>_{q_{\bar{k}}} \nonumber \\
	&\qquad + (\eta-1)\ln \gamma_{k,l} - \gamma_{k,l} \nu
\label{eq:AppEq4}
\end{align}
where $q_{\bar{k}}$ is the product of all the factors $q_k \in \mathcal{Q} \backslash \{q_{\gamma,k,l}\}$.
Since $\bm{\Gamma}=\text{diag}(\bm{\gamma}_{\hat{\mathcal{S}}})$, we can simplify $\ln |\pi\bm{\Gamma}^{-1}|^{-1}=\ln \gamma_{k,l}+\text{const.}$ and $\bm{\alpha}_{\hat{\mathcal{S}}}^\hermitian\bm{\Gamma}\bm{\alpha}_{\hat{\mathcal{S}}}=\gamma_{k,l}|\alpha_{k,l}|^2 + \text{const.}$
Thus, \eqref{eq:AppEq4} can be rewritten as a Gamma \gls{pdf}, i.e.,  
\begin{equation}
	\ln p_{\gamma,k,l} \overset{e}{\propto} \eta \ln \gamma_{k,l} - \gamma_{k,l}(|\hat{\alpha}_{k,l}^2| + \hat{C}_{k,l} + \nu)
\end{equation}
with shape $\eta+1$ and rate $|\hat{\alpha}_{k,l}|^2+\hat{C}_{k,l}+\nu$ as in \eqref{eq:q-gamma} and \eqref{eq:update_gamma_single}.
The derivation for $\hat{z}_k=0$ is omitted because we consider it to be trivial.

\subsection{Derivation of $\ln Z(\theta,z)$ and $\Delta_k(\theta_k)$}
\label{sec:appendix-lnZ-maximization}

We start by showing that expression $\mathrm{I}$ in \eqref{eq:helperPDF} is the logarithm of a complex Gaussian distribution for $\bm{\alpha}_{\tilde{\mathcal{S}}}$ and $\alpha_{k,l}=0$ for $\alpha_{k,l}\notin \bm{\alpha}_{\tilde{\mathcal{S}}}$ plus an expressions which depends on $\bm{\theta}$ and $\bm{z}$.
Let $\bm{\Gamma}_{\tilde{\mathcal{S}}}=\text{diag}(\bm{\gamma}_{\tilde{\mathcal{S}}})$,
\begin{align}
	\mathrm{I} &=\big<\ln p(\bm{\alpha},\bm{\theta},\bm{\gamma},\bm{z},\lambda|\bm{x})\big>_{q_{\lambda}q_{\bm{\gamma}}} \nonumber \\
	&\overset{e}{\propto} \big<-\lambda(\bm{x}-\bm{\Psi}_{\tilde{\mathcal{S}}}\bm{\alpha}_{\tilde{\mathcal{S}}})^\hermitian(\bm{x}-\bm{\Psi}_{\tilde{\mathcal{S}}}\bm{\alpha}_{\tilde{\mathcal{S}}})+\ln |\pi\bm{\Gamma}_{\tilde{S}}|^{-1} \nonumber \\
	&\qquad  -\bm{\alpha}_{\tilde{\mathcal{S}}}^\hermitian\bm{\Gamma}_{\tilde{\mathcal{S}}}\bm{\alpha}_{\tilde{\mathcal{S}}} +\ln p(\bm{z}) + \sum_{\alpha_{k,l}\notin\bm{\alpha}_{\tilde{\mathcal{S}}}}\ln \delta(\alpha_{k,l}) \big>_{q_{\lambda}q_{\bm{\gamma}}}.\\[-7mm]\nonumber
\end{align}
After ``completing the squares'' and adding $\ln |\pi \bm{C}_{\tilde{\mathcal{S}}}|^{-1}-\ln |\pi \bm{C}_{\tilde{\mathcal{S}}}|^{-1}=0$ to complete the Gaussian distribution we find
\begin{align}
\hspace*{-1.0mm} \mathrm{I} &\overset{e}{\propto}
\ln |\pi \bm{C}_{\tilde{\mathcal{S}}}|^{-1} -(\bm{\alpha}_{\tilde{\mathcal{S}}}-\hat{\bm{\alpha}})^\hermitian\bm{C}_{\tilde{\mathcal{S}}}^{-1}(\bm{\alpha}_{\tilde{\mathcal{S}}}-\hat{\bm{\alpha}})  \nonumber \\
&\hspace*{1.4mm} + \sum_{\alpha_{k,l}\notin \bm{\alpha}_{\tilde{\mathcal{S}}}} \ln \delta(\alpha_{k,l}) + \hat{\lambda}^2\bm{x}^\hermitian \bm{\Psi}_{\tilde{\mathcal{S}}}\bm{C}_{\tilde{\mathcal{S}}}\bm{\Psi}_{\tilde{\mathcal{S}}}^\hermitian\bm{x} -\ln |\pi \bm{C}_{\tilde{\mathcal{S}}}|^{-1}   \nonumber \\
&\hspace*{1.4mm} + \big<\ln |\pi\bm{\Gamma}_{\tilde{\mathcal{S}}}|^{-1}\big>_{q_{\bm{\gamma}}} + \ln p(\bm{z})\nonumber\\
&\overset{e}{\propto}
\ln \mathcal{CN}(\bm{\alpha}_{\tilde{\mathcal{S}}}|\hat{\bm{\alpha}},\bm{C}_{\tilde{\mathcal{S}}})+\sum_{\alpha_{k,l}\notin \bm{\alpha}_{\tilde{\mathcal{S}}}} \ln \delta(\alpha_{k,l}) + \ln p(\bm{z}) \nonumber \\
&\hspace*{1.4mm} + \hat{\lambda}^2\bm{x}^\hermitian \bm{\Psi}_{\tilde{\mathcal{S}}}\bm{C}_{\tilde{\mathcal{S}}}\bm{\Psi}_{\tilde{\mathcal{S}}}^\hermitian\bm{x} - \ln |\bm{C}_{\tilde{\mathcal{S}}}| + \hspace*{-1.4mm}\sum_{\gamma_{k,l}\in\bm{\gamma}_{\tilde{\mathcal{S}}}}\hspace*{-1.2mm}\big<\ln \gamma_{k,l}\big>_{q_{\gamma,k,l}}.
\label{eq:appendix:lnZ-itegral}\\[-7mm]\nonumber
\end{align}
Inserting \eqref{eq:appendix:lnZ-itegral} into \eqref{eq:activations_distribution_t}, all terms which depend on $\bm{\alpha}$ integrate to 1 since they form a valid distribution.
Thus, after integrating out $\bm{\alpha}$ and taking the logarithm we arrive at \eqref{eq:elbo_maximization_global}.

Let $\bm{C}_k = (\hat{\lambda}\bm{\Psi}_k^\hermitian\bm{\Psi}_k + \bm{\Gamma}_k - \hat{\lambda}^{2}\bm{\Psi}_k\bm{\Psi}_{k}^\hermitian\bm{\Psi}_{\bar{k}}\bm{C}_{\bar{k}}\bm{\Psi}_{\bar{k}}^\hermitian\bm{\Psi}_k)^{-1}$, we express the covariance matrix $\bm{C}_{\tilde{\mathcal{S}}}$ in $\ln Z([\hat{\bm{\theta}}_{\bar{k}},\,\theta_k],[\hat{\bm{z}}_{\bar{k}},1])$ as a block matrix
\begin{equation}
	\bm{C}_{\tilde{\mathcal{S}}} = \left[\begin{matrix}
		\bm{C}_{\bar{k}}^{-1} & \hat{\lambda}\bm{\Psi}_{\bar{k}}^\hermitian\bm{\Psi}_{k} \\
		\hat{\lambda}\bm{\Psi}_{k}^\hermitian\bm{\Psi}_{\bar{k}} & (\hat{\lambda}\bm{\Psi}_k^\hermitian\bm{\Psi}_k + \bm{\Gamma}_k)
	\end{matrix}\right]^{-1}
\end{equation}
and use the formula for block-matrix inversion to find $\ln |\bm{C}_{\tilde{\mathcal{S}}}| = \ln |\bm{C}_{\bar{k}}| + \ln |\bm{C}_k|$ and
\begin{align}
\hspace*{-0.4mm}&\hat{\lambda}^2\bm{x}^\hermitian\bm{\Psi}_{\tilde{\mathcal{S}}}\bm{C}_{\tilde{\mathcal{S}}}\bm{\Psi}_{\tilde{\mathcal{S}}}^\hermitian\bm{x}\nonumber\\
&=\hat{\lambda}^4\bm{x}^\hermitian\bm{\Psi}_{\bar{k}}\bm{C}_{\bar{k}}\bm{\Psi}_{\bar{k}}^\hermitian\bm{\Psi}_k\bm{C}_k\bm{\Psi}_k^\hermitian\bm{\Psi}_{\bar{k}}\bm{C}_{\bar{k}}\bm{\Psi}_{\bar{k}}^\hermitian\bm{x} + \hat{\lambda}^2\bm{x}^\hermitian\bm{\Psi}_{\bar{k}}\bm{C}_{\bar{k}}\bm{\Psi}_{\bar{k}}^\hermitian\bm{x}\nonumber \\
&\hspace*{2mm} +\hspace*{-0.3mm}\hat{\lambda}^2\bm{x}^\hermitian\bm{\Psi}_k\bm{C}_k\bm{\Psi}_k^\hermitian\bm{x}  \hspace*{-0.2mm}-\hspace*{-0.2mm}\Re\{2\hat{\lambda}^3\bm{x}^\hermitian\bm{\Psi}_k\bm{C}_k\bm{\Psi}_k^\hermitian\bm{\Psi}_{\bar{k}}\bm{C}_{\bar{k}}\bm{\Psi}_{\bar{k}}^\hermitian\bm{x}\}\nonumber\\[-7mm]\nonumber
\end{align}
which simplifies to
\begin{align}
	\hat{\lambda}^2\bm{x}^\hermitian \bm{\Psi}_{\tilde{\mathcal{S}}}\bm{C}_{\tilde{\mathcal{S}}}\bm{\Psi}_{\tilde{\mathcal{S}}}^\hermitian\bm{x} = \hat{\lambda}^2\bm{x}^\hermitian\bm{\Psi}_{\bar{k}}\bm{C}_{\bar{k}}\bm{\Psi}_{\bar{k}}^\hermitian\bm{x} + \bm{u}^\hermitian\bm{C}_k^{-1}\bm{u}
	\label{eq:appendix:block-matrix-expansion}.\\[-7mm]\nonumber
\end{align}
Note, that $\bm{C}_{\tilde{\mathcal{S}}}$ in the calculation of $\ln Z([\hat{\bm{\theta}}_{\bar{k}},\,\theta_k],[\hat{\bm{z}}_{\bar{k}},0])$ equals $\bm{C}_{\bar{k}}$ in the calculation of $\ln Z([\hat{\bm{\theta}}_{\bar{k}},\,\theta_k],[\hat{\bm{z}}_{\bar{k}},1])$.
The same holds for $\bm{\Psi}_{\tilde{\mathcal{S}}}$ and $\bm{\Psi}_{\bar{k}}$, respectively.
Thus, we insert \eqref{eq:appendix:block-matrix-expansion} into \eqref{eq:elbo_maximization_global} and arrive at \eqref{eq:elbo_difference_group} after a few algebraic manipulations.

\bibliography{IEEEabrv,references_moederl2023TSP}

\begin{thebibliography}{10}
\providecommand{\url}[1]{#1}
\csname url@samestyle\endcsname
\providecommand{\newblock}{\relax}
\providecommand{\bibinfo}[2]{#2}
\providecommand{\BIBentrySTDinterwordspacing}{\spaceskip=0pt\relax}
\providecommand{\BIBentryALTinterwordstretchfactor}{4}
\providecommand{\BIBentryALTinterwordspacing}{\spaceskip=\fontdimen2\font plus
\BIBentryALTinterwordstretchfactor\fontdimen3\font minus
  \fontdimen4\font\relax}
\providecommand{\BIBforeignlanguage}[2]{{%
\expandafter\ifx\csname l@#1\endcsname\relax
\typeout{** WARNING: IEEEtran.bst: No hyphenation pattern has been}%
\typeout{** loaded for the language `#1'. Using the pattern for}%
\typeout{** the default language instead.}%
\else
\language=\csname l@#1\endcsname
\fi
#2}}
\providecommand{\BIBdecl}{\relax}
\BIBdecl

\bibitem{stoica2005:SpectralAnalysis}
P.~Stoica and R.~Moses, \emph{\BIBforeignlanguage{eng}{Spectral analysis of
  signals}}.\hskip 1em plus 0.5em minus 0.4em\relax Upper Saddle River, NJ,
  USA: Pearson Prentice Hall, 2005.

\bibitem{benents2019SPM}
E.~Benetos, S.~Dixon, Z.~Duan, and S.~Ewert, ``Automatic music transcription:
  An overview,'' \emph{{IEEE} Signal Process. Mag.}, vol.~36, no.~1, pp.
  20--30, Jan. 2019.

\bibitem{mueller2011STSP:musicAnalysis}
M.~Müller, D.~P.~W. Ellis, A.~Klapuri, and G.~Richard, ``Signal processing for
  music analysis,'' \emph{{IEEE} J. Sel. Topics Signal Process.}, vol.~5,
  no.~6, pp. 1088--1110, Feb. 2011.

\bibitem{christensen2009:MultiPitchEst}
M.~G. Christensen and A.~Jakobsson, \emph{Multi-Pitch Estimation}, ser.
  Synthesis Lectures on Speech \& Audio Processing, B.~H. Juang, Ed.\hskip 1em
  plus 0.5em minus 0.4em\relax San Rafael, CA, USA: Morgan \& Claypool, 2009.

\bibitem{christensen2008SP}
M.~G. Christensen, P.~Stoica, A.~Jakobsson, and S.~{Holdt Jensen},
  ``Multi-pitch estimation,'' \emph{Signal Process.}, vol.~88, no.~4, pp.
  972--983, Apr. 2008.

\bibitem{granstrom2017JAIF:EO}
K.~Granstrom, M.~Baum, and S.~Reuter, ``Extended object tracking: Introduction,
  overview, and applications,'' \emph{J. Advances Inf. Fusion}, vol.~12, no.~2,
  pp. 139--174, Dec. 2017.

\bibitem{schubert2013TAP}
F.~M. Schubert, M.~L. Jakobsen, and B.~H. Fleury, ``Non-stationary propagation
  model for scattering volumes with an application to the rural {LMS}
  channel,'' \emph{{IEEE} Trans. Antennas Propag.}, vol.~61, no.~5, pp.
  2817--2828, Jan. 2013.

\bibitem{dragomiretskiy2014TSP:VMD}
K.~Dragomiretskiy and D.~Zosso, ``Variational mode decomposition,''
  \emph{{IEEE} Trans. Signal Process.}, vol.~62, no.~3, pp. 531--544, Feb.
  2014.

\bibitem{schmidt1986TAP}
R.~Schmidt, ``Multiple emitter location and signal parameter estimation,''
  \emph{{IEEE} Trans. Antennas Propag.}, vol.~34, no.~3, pp. 276--280, Mar.
  1986.

\bibitem{roy1989TASSP}
R.~Roy and T.~Kailath, ``{ESPRIT}-estimation of signal parameters via
  rotational invariance techniques,'' \emph{{IEEE} Trans. Acoust., Speech,
  Signal Process.}, vol.~37, no.~7, pp. 984--995, Jul. 1989.

\bibitem{ziskind1988TASSP:ML}
I.~Ziskind and M.~Wax, ``Maximum likelihood localization of multiple sources by
  alternating projection,'' \emph{{IEEE} Trans. Acoust., Speech, Signal
  Process.}, vol.~36, no.~10, pp. 1553--1560, Oct. 1988.

\bibitem{feder1988TASSP:ML}
M.~Feder and E.~Weinstein, ``Parameter estimation of superimposed signals using
  the {EM} algorithm,'' \emph{{IEEE} Trans. Acoust., Speech, Signal Process.},
  vol.~36, no.~4, pp. 477--489, Apr. 1988.

\bibitem{stoica2004SPM}
P.~Stoica and Y.~Selen, ``Model-order selection: a review of information
  criterion rules,'' \emph{{IEEE} Signal Process. Mag.}, vol.~21, no.~4, pp.
  36--47, Jul. 2004.

\bibitem{tibshirani1996RSSB}
R.~Tibshirani, ``Regression shrinkage and selection via the {LASSO},'' \emph{J.
  Roy. Statistical Soc.: Ser. B (Statistical Methodology)}, vol.~58, no.~1, pp.
  267--288, 1996.

\bibitem{chen2001SIAMRev}
S.~S. Chen, D.~L. Donoho, and M.~A. Saunders, ``Atomic decomposition by basis
  pursuit,'' \emph{{SIAM} Rev.}, vol.~43, no.~1, pp. 129--159, Mar. 2001.

\bibitem{mallat1993TSP:MP}
S.~Mallat and Z.~Zhang, ``Matching pursuits with time-frequency dictionaries,''
  \emph{{IEEE} Trans. Signal Process.}, vol.~41, no.~12, pp. 3397--3415, 1993.

\bibitem{tipping1999NeurIPS:RelevanceVector}
M.~Tipping, ``The relevance vector machine,'' in \emph{Advances Neural Inf.
  Process. Syst.}, vol.~12.\hskip 1em plus 0.5em minus 0.4em\relax Denver, CO,
  USA: MIT Press, Nov. 29 -- Dec. 4, 1999, pp. 652--658.

\bibitem{tipping2003WAIS:FastMarginalSparseBayesian}
M.~E. Tipping and A.~C. Faul, ``Fast marginal likelihood maximisation for
  sparse {Bayesian} models,'' in \emph{Proc. 9th Int. Workshop Artif. Intell.
  and Statist.}, vol.~R4, Key West, FL, USA, Jan. 03--06, 2003, pp. 276--283.

\bibitem{shutin2011TSP:fastVSBL}
D.~Shutin, T.~Buchgraber, S.~R. Kulkarni, and H.~V. Poor, ``Fast variational
  sparse {Bayesian} learning with automatic relevance determination for
  superimposed signals,'' \emph{{IEEE} Trans. Signal Process.}, vol.~59,
  no.~12, pp. 6257--6261, Dec. 2011.

\bibitem{stoica2011TSP:SPICE}
P.~Stoica, P.~Babu, and J.~Li, ``{SPICE}: A sparse covariance-based estimation
  method for array processing,'' \emph{{IEEE} Trans. Signal Process.}, vol.~59,
  no.~2, pp. 629--638, Feb. 2011.

\bibitem{wipf2004TSP}
D.~P. Wipf and B.~D. Rao, ``Sparse bayesian learning for basis selection,''
  \emph{{IEEE} Trans. Signal Process.}, vol.~52, no.~8, pp. 2153--2164, Aug.
  2004.

\bibitem{wipf2011TIP}
D.~P. Wipf, B.~D. Rao, and S.~Nagarajan, ``Latent variable {Bayesian} models
  for promoting sparsity,'' \emph{{IEEE} Trans. Image Process.}, vol.~57,
  no.~9, pp. 6236--6255, Sep. 2011.

\bibitem{yuan2006RSSB}
M.~Yuan and Y.~Lin, ``Model selection and estimation in regression with grouped
  variables,'' \emph{J. Roy. Statistical Soc.: Ser. B (Statistical
  Methodology)}, vol.~68, no.~1, pp. 49--67, Feb. 2006.

\bibitem{kyung2010BA}
M.~Kyung, J.~Gill, M.~Ghosh, and G.~Casella, ``Penalized regression, standard
  errors, and {Bayesian} {LASSO}s,'' \emph{Bayesian Anal.}, vol.~5, no.~2, pp.
  369--411, Jun. 2010.

\bibitem{raman2009ICML}
S.~Raman, T.~J. Fuchs, P.~J. Wild, E.~Dahl, and V.~Roth, ``The {Bayesian}
  group-{LASSO} for analyzing contingency tables,'' in \emph{Proc. 26th Annu.
  Int. Conf. Mach. Learn.}, New York, NY, USA, Jun. 14--18, 2009, pp. 881--888.

\bibitem{xu2015BA}
X.~Xu and M.~Ghosh, ``{Bayesian} variable selection and estimation for group
  {LASSO},'' \emph{Bayesian Anal.}, vol.~10, no.~4, pp. 909--936, Dec. 2015.

\bibitem{kim2006SS:BSR}
Y.~Kim, J.~Kim, and Y.~Kim, ``Blockwise sparse regression,'' \emph{Statistica
  Sinica}, vol.~16, no.~2, pp. 375--390, Apr. 2006.

\bibitem{eldar2010TSP:BlockSparseMP}
Y.~C. Eldar, P.~Kuppinger, and H.~Bolcskei, ``Block-sparse signals: Uncertainty
  relations and efficient recovery,'' \emph{{IEEE} Trans. Signal Process.},
  vol.~58, no.~6, pp. 3042--3054, Jun. 2010.

\bibitem{zhang2011STSP:bSBL}
Z.~Zhang and B.~D. Rao, ``Sparse signal recovery with temporally correlated
  source vectors using sparse bayesian learning,'' \emph{{IEEE} J. Sel. Topics
  Signal Process.}, vol.~5, no.~5, pp. 912--926, 2011.

\bibitem{zhang2013TSP:BlockSparseSBL}
------, ``Extension of {SBL} algorithms for the recovery of block sparse
  signals with intra-block correlation,'' \emph{{IEEE} Trans. Signal Process.},
  vol.~61, no.~8, pp. 2009--2015, Apr. 2013.

\bibitem{fang2015TSP:PatternCoupledSBL}
J.~Fang, Y.~Shen, H.~Li, and P.~Wang, ``Pattern-coupled sparse bayesian
  learning for recovery of block-sparse signals,'' \emph{{IEEE} Trans. Signal
  Process.}, vol.~63, no.~2, pp. 360--372, Jan. 2015.

\bibitem{kronvall2017SP:groupSparseRegression}
T.~Kronvall, S.~I. Adalbjörnsson, S.~Nadig, and A.~Jakobsson, ``Group-sparse
  regression using the covariance fitting criterion,'' \emph{Signal Process.},
  vol. 139, pp. 116--130, Oct. 2017.

\bibitem{chi2011TSP}
Y.~Chi, L.~L. Scharf, A.~Pezeshki, and A.~R. Calderbank, ``Sensitivity to basis
  mismatch in compressed sensing,'' \emph{{IEEE} Trans. Signal Process.},
  vol.~59, no.~5, pp. 2182--2195, May 2011.

\bibitem{duarte2013ACHA}
M.~F. Duarte and R.~G. Baraniuk, ``Spectral compressive sensing,'' \emph{Appl.
  Comput. Harmon. Anal.}, vol.~35, no.~1, pp. 111--129, Jul. 2013.

\bibitem{yang2015TSP:GLS}
Z.~Yang and L.~Xie, ``On gridless sparse methods for line spectral estimation
  from complete and incomplete data,'' \emph{{IEEE} Trans. Signal Process.},
  vol.~63, no.~12, pp. 3139--3153, Jun. 2015.

\bibitem{hansen2018TSP:SuperFastLSE}
T.~L. Hansen, B.~H. Fleury, and B.~D. Rao, ``Superfast line spectral
  estimation,'' \emph{{IEEE} Trans. Signal Process.}, vol.~66, no.~10, pp.
  2511--2526, Feb. 2018.

\bibitem{hansen2014SAM:SBL}
T.~L. Hansen, M.~A. Badiu, B.~H. Fleury, and B.~D. Rao, ``A sparse {Bayesian}
  learning algorithm with dictionary parameter estimation,'' in \emph{2014 IEEE
  8th Sensor Array and Multichannel Signal Process. Workshop (SAM)}, A Coruna,
  Spain, Jun. 22--25, 2014, pp. 385--388.

\bibitem{shutin2013:VSBL}
D.~Shutin, W.~Wand, and T.~Jost, ``Incremental sparse {Bayesian} learning for
  parameter estimation of superimposed signals,'' in \emph{10th Int. Conf.
  Sampling Theory and Appl.}, Bremen, Germany, Jul. 1--5, 2013, pp. 513--516.

\bibitem{badiu2017TSP:VSBL}
M.-A. Badiu, T.~L. Hansen, and B.~H. Fleury, ``Variational {Bayesian} inference
  of line spectra,'' \emph{{IEEE} Trans. Signal Process.}, vol.~65, no.~9, pp.
  2247--2261, May 2017.

\bibitem{pedersen2015SP}
N.~L. Pedersen, C.~{Navarro Manchón}, M.-A. Badiu, D.~Shutin, and B.~H.
  Fleury, ``Sparse estimation using {Bayesian} hierarchical prior modeling for
  real and complex linear models,'' \emph{Signal Process.}, vol. 115, pp.
  94--109, Oct. 2015.

\bibitem{adalbjoernsson2015SP}
S.~I. Adalbjörnsson, A.~Jakobsson, and M.~G. Christensen, ``Multi-pitch
  estimation exploiting block sparsity,'' \emph{Signal Process.}, vol. 109, pp.
  236--247, Apr. 2015.

\bibitem{swaerd2018TASLP}
J.~Swärd, H.~Li, and A.~Jakobsson, ``Off-grid fundamental frequency
  estimation,'' \emph{{IEEE/ACM} Trans. Audio, Speech, Language Process.},
  vol.~26, no.~2, pp. 296--303, Feb. 2018.

\bibitem{vincent2008NC:BayesianHarmonic}
E.~Vincent and M.~D. Plumbley, ``Efficient bayesian inference for harmonic
  models via adaptive posterior factorization,'' \emph{Neurocomput.}, vol.~72,
  no.~1, pp. 79--87, Dec. 2008.

\bibitem{tzikas2008:VAEM}
D.~G. Tzikas, A.~C. Likas, and N.~P. Galatsanos, ``The variational
  approximation for {Bayesian} inference,'' \emph{{IEEE} Signal Process. Mag.},
  vol.~25, no.~6, pp. 131--146, Nov. 2008.

\bibitem{Bishop2006}
C.~M. Bishop, \emph{Pattern Recognition and Machine Learning (Information
  Science and Statistics)}.\hskip 1em plus 0.5em minus 0.4em\relax Secaucus,
  NJ, USA: Springer-Verlag New York, Inc., 2006.

\bibitem{leitinger2020Asilomar}
E.~Leitinger, S.~Grebien, B.~Fleury, and K.~Witrisal, ``Detection and
  estimation of a spectral line in {MIMO} systems,'' in \emph{2020 54th
  Asilomar Conf. Signals, Syst. and Computers}, Pacific Grove, CA, USA, Nov.
  01--04, 2020, pp. 1090--1095.

\bibitem{marple1999TSP:analytic-signal}
L.~Marple, ``Computing the discrete-time ``analytic'' signal via {FFT},''
  \emph{{IEEE} Trans. Signal Process.}, vol.~47, no.~9, pp. 2600--2603, Sep.
  1999.

\bibitem{schuhmacher2008TSP:OSPA}
D.~Schuhmacher, B.-T. Vo, and B.-N. Vo, ``A consistent metric for performance
  evaluation of multi-object filters,'' \emph{{IEEE} Trans. Signal Process.},
  vol.~56, no.~8, pp. 3447--3457, Aug. 2008.

\bibitem{duan2010TASLP:Bach10}
Z.~Duan, B.~Pardo, and C.~Zhang, ``Multiple fundamental frequency estimation by
  modeling spectral peaks and non-peak regions,'' \emph{{IEEE/ACM} Trans.
  Audio, Speech, Language Process.}, vol.~18, no.~8, pp. 2121--2133, Nov. 2010.

\bibitem{cheveigne2002JASA:YIN}
A.~de~Cheveigné and H.~Kawahara, ``{YIN}, a fundamental frequency estimator
  for speech and music,'' \emph{J. Acoust. Soc. Amer.}, vol. 111, no.~4, pp.
  1917--1930, Apr. 2022.

\bibitem{bay2009ISMIR}
M.~Bay, A.~F. Ehmann, and J.~S. Downie, ``Evaluation of multiple-{F0}
  estimation and tracking systems,'' in \emph{Proc. 10th Int. Soc. Music Inf.
  Retrieval Conf.}, Kobe, Japan, Oct. 26 -- 30, 2009, pp. 315--320.

\bibitem{elvander2017TASLP}
F.~Elvander, J.~Swärd, and A.~Jakobsson, ``Online estimation of multiple
  harmonic signals,'' \emph{{IEEE/ACM} Trans. Audio, Speech, Language
  Process.}, vol.~25, no.~2, pp. 273--284, Feb. 2017.

\bibitem{benetos2015ISMIR}
E.~Benetos and T.~Weyde, ``An efficient temporally-constrained probabilistic
  model for multiple-instrument music transcription,'' in \emph{Proc. 16th Int.
  Soc. Music Inf. Retrieval Conf.}, Malaga, Spain, Oct. 26--30, 2015.

\bibitem{li2022CC}
X.~Li, Y.~Yan, J.~Soraghan, Z.~Wang, and J.~Ren, ``A music cognition--guided
  framework for multi-pitch estimation,'' \emph{Cogn. Comput.}, pp. 1--13,
  2022.

\bibitem{xuhong2022TWC}
X.~Li, E.~Leitinger, A.~Venus, and F.~Tufvesson, ``Sequential detection and
  estimation of multipath channel parameters using belief propagation,''
  \emph{{IEEE} Trans. Wireless Commun.}, vol.~21, no.~10, pp. 8385--8402, Oct.
  2022.

\bibitem{LeitMeyHlaWitTufWin:TWC2019}
E.~{Leitinger}, F.~{Meyer}, F.~{Hlawatsch}, K.~{Witrisal}, F.~{Tufvesson}, and
  M.~Z. {Win}, ``A belief propagation algorithm for multipath-based {SLAM},''
  \emph{{IEEE} Trans. Wireless Commun.}, vol.~18, no.~12, pp. 5613--5629, Dec.
  2019.

\bibitem{LeiVenTeaMey:Arxiv2022}
\BIBentryALTinterwordspacing
E.~{Leitinger}, A.~{Venus}, B.~{Teague}, and F.~{Meyer}, ``Data fusion for
  multipath-based {SLAM}: {Combining} information from multiple propagation
  paths,'' \emph{ArXiv e-prints}, 2022. [Online]. Available:
  \url{https://arxiv.org/abs/2211.09241}
\BIBentrySTDinterwordspacing

\bibitem{johnson2016NeurIPS:SVAE}
M.~J. Johnson, D.~K. Duvenaud, A.~Wiltschko, R.~P. Adams, and S.~R. Datta,
  ``Composing graphical models with neural networks for structured
  representations and fast inference,'' in \emph{Advances Neural Inf. Process.
  Syst.}, vol.~29.\hskip 1em plus 0.5em minus 0.4em\relax Barcelona, Spain:
  Curran Associates, Inc., Dec. 5--10, 2016.

\end{thebibliography}
\bibliographystyle{IEEEtran}

\end{document}